\documentclass[cernpreprint, atlasdraft=false, UKenglish, texmf, orcidlogo]{atlasdoc}
\usepackage[subcaption]{atlaspackage}
\usepackage{atlasbiblatex}

\usepackage{lscape}

\usepackage{atlasphysics}
\graphicspath{{logos/}{figures/}}

\usepackage{ANA-HIGP-2024-05-PAPER-defs}
\usepackage{array}
\usepackage{multirow}
\usepackage{pifont}
\usepackage{booktabs}

\usepackage{tikz}
\usepackage[compat=1.1.0]{tikz-feynman}
\definecolor{coolgrey}{rgb}{0.55, 0.57, 0.67}
\definecolor{amaranth}{rgb}{0.9, 0.17, 0.31}
\definecolor{burgundy}{rgb}{0.5, 0.0, 0.13}

\AtlasTitle{Constraining off-shell Higgs boson production and the Higgs boson total width using $WW\to \ell\nu\ell\nu$ final states with the ATLAS detector}

\AtlasAbstract{
A measurement of off-shell Higgs boson production is performed in the $H^{*} \rightarrow WW$ channel. The measurement uses a proton--proton collision dataset with an integrated luminosity of 140 fb$^{-1}$ collected at a centre-of-mass energy of 13 TeV by the ATLAS detector at the Large Hadron Collider. Final states in which both $W$~bosons decay leptonically are targeted, and events are categorised based on the flavour of the final-state leptons, the jet multiplicity, and the output of neural network-based classifiers. The data are found to be compatible with the Standard Model expectation. An observed (expected) upper bound on the 95\% symmetric confidence level interval is set on the rate of off-shell Higgs boson production at a value of 3.4 (4.4) times the Standard Model prediction. These results are combined with the results from the measurement of on-shell Higgs boson production in the same final states to obtain an observed (expected) upper bound at 95$\%$ confidence level on the Higgs boson total width of 13.1 (17.3) MeV.}

\PreprintIdNumber{CERN-EP-2025-059}

\AtlasJournal{Phys.\ Lett.\ B.}
\AtlasJournalRef{Phys.\ Lett.\ B 870 (2025) 139898}
\AtlasDOI{10.1016/j.physletb.2025.139898}

\hypersetup{pdftitle={ATLAS document},pdfauthor={The ATLAS Collaboration}}

\begin{document}

\maketitle

\section{Introduction}
\label{sec:intro}
%

% The next lines are included from the .//sections/intro.tex input file
Since its discovery in the year 2012 by the ATLAS~\cite{HIGG-2012-27} and CMS~\cite{CMS-HIG-12-028} Collaborations, the Higgs boson has been at the centre of a major campaign of searches and measurements. These efforts aim to test the compatibility of the Higgs boson's properties with the Standard Model (SM) predictions.
No significant discrepancies have been found to date, but the current precision of the measurements leaves substantial room for beyond-the-SM (BSM) effects for several parameters, for example the total width of the Higgs boson, $\Gamma_H$.

As $\Gamma_{H}$ represents the sum of the partial widths of the individual decay modes of the Higgs, any modification caused by an additional BSM interaction would impact its value. Values of the Higgs boson width either above or below the SM prediction occur in various formulations of BSM physics, including composite Higgs models \cite{agashe2005minimal}, two-Higgs-doublet models \cite{branco2012theory}, and various scalar extensions of the SM~\cite{Cao:2023}. Therefore, a precise measurement of $\Gamma_{H}$ may provide sensitivity to BSM physics.

In the SM, the Higgs boson width can be computed using the known masses and couplings of all particles to which the Higgs boson decays, which yields a value of $4.1$\,\MeV with an associated uncertainty of approximately 1\%~\cite{deFlorian:2016spz}. Due to the limited resolution of the ATLAS and CMS detectors, a direct measurement of the width from the Higgs boson mass peak~\cite{CMS-HIG-21-019,HIGG-2020-07} cannot reach the \MeV-scale precision needed to probe this quantity experimentally. However, this level of precision can be achieved through an indirect method proposed in Refs.~\cite{Passarino:2012ri,Campbell:2014,Caola:2013}, which uses Higgs boson decays to pairs of vector bosons $VV$, where $VV=WW$ or $ZZ$. This method compares the rates of Higgs boson production measured across two parts of the phase space distinguished by the invariant mass of the vector boson pair, $m_{VV}$: when $m_{VV}$ is close to the Higgs boson pole mass, $m_{H}$ (on-shell regime), and when $m_{VV}$ is much greater than that value (off-shell regime).\footnote{In this analysis the threshold between on-shell and off-shell production is set to $m_{VV} = 140$\,\GeV as defined in Ref.~\cite{deFlorian:2016spz}.} The rate of $VV$ production mediated by an on-shell Higgs boson $H$~depends on the Higgs boson's couplings to particles in its production and decay, $g_{\mathrm{prod}}(m_H)$ and $g_{\mathrm{decay}}(m_H)$, as well as its total width, $\Gamma_H$. In contrast, the rate of $VV$ production mediated by an off-shell Higgs boson $H^{*}$~is independent of the total width but still depends on the couplings $g_{\mathrm{prod}}(\hat{s})$ and $g_{\mathrm{decay}}(\hat{s})$, where  $\hat{s}$  represents the virtuality of the Higgs boson.

Deviations from the SM predictions for these rates can be expressed in terms of the on-shell and off-shell signal strengths, $\muonshell$ and $\muoff$, as follows:
\begin{equation}
\muonshell= \kappa_{\mathrm{prod, on\mbox{-}shell}}^2 \cdot \kappa_{\mathrm{decay, on\mbox{-}shell}}^2 \cdot \frac{1}{\kappa_H};
~~~~~~~ \muoff=\kappa_{\mathrm{prod, off\mbox{-}shell}}^2 \cdot \kappa_{\mathrm{decay, off\mbox{-}shell}}^2,
\end{equation}
where $\kappa_H=\Gamma_{H}/\Gamma_{H}^\mathrm{SM}$, and the other $\kappa$s represent coupling strength modifiers with respect to the SM predictions, $\kappa_{i} = g_{i}/ g_{i}^{\mathrm{SM}}$, where $i=\mathrm{prod}$ or decay. Assuming that the coupling strength modifiers are the same for both regimes, $R_{i} = \kappa_{i,\mathrm{off\mbox{-}shell}}^2 / \kappa_{i,\mathrm{on\mbox{-}shell}}^2 = 1$, the total Higgs boson width can be extracted by taking the ratio of measured on-shell and off-shell signal strengths.\footnote{This assumption may be violated in some BSM theories, including the Georgi-Machacek (GM) model \cite{georgi1985doubly} and particular Sigma models \cite{alonso2016sigma}.} Experimentally, this method is feasible because off-shell production in \hstarvv~channels is enhanced for invariant masses $m_{VV}$ above $2m_V$, as the vector bosons themselves become on-shell. This effect partially compensates the rapid decrease in the rate of Higgs boson production that occurs away from its mass peak.

\begin{figure}
\centering
\subfloat[]{
\includegraphics[width=.4\textwidth]{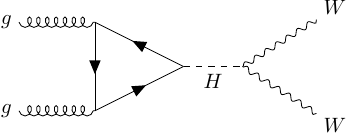}
}
\hspace{0.05\textwidth}
\subfloat[]{
\includegraphics[width=.4\textwidth]{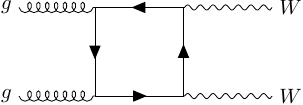}
}\\
\hspace{0.01\textwidth}
\subfloat[]{
\includegraphics[width=.35\textwidth]{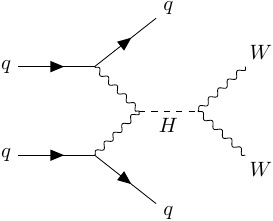}
}
\hspace{0.11\textwidth}
\subfloat[]{
\includegraphics[width=.35\textwidth]{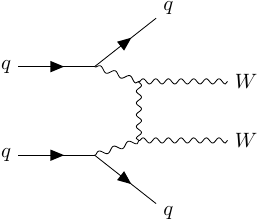}
}
\caption{Example diagrams of leading-order gluon-induced $WW$ production (ggF) (a) including and (b) not including a Higgs boson, and quark-induced $WW$ production (EW) (c) including and (d) not including a Higgs boson.\label{fig:intro:feynmandiagrams}}
\end{figure}

Evidence for off-shell Higgs boson production in the \hstarzz~channel has been reported by both the CMS and ATLAS Collaborations, using the full LHC Run~2 data set in the $4\ell$ and $2\ell2\nu$ final states, where $\ell$ denotes an electron or muon. The ATLAS Collaboration published an analysis employing neural simulation-based inference~\cite{HIGP-2024-14}, yielding $\Gamma_H = 4.3^{+2.7}_{-1.9}~\MeV$ with a significance of $3.7\sigma$. The CMS Collaboration presented a significance of $3.8\sigma$, corresponding to a constraint of $\Gamma_H = 3.0^{+2.0}_{-1.5}~\MeV$~\cite{CMS-HIG-21-019, CMS-HIG-21-013}. Another strategy to measure $\Gamma_H$, which relies on the measurement of the production of four top quarks, was recently performed by the ATLAS Collaboration leading to an upper limit on $\Gamma_H$ of 450\,\MeV~\cite{TOPQ-2023-22}. The latest published results from the \hstarww~channel are from 2015~\cite{HIGG-2014-10} and 2016~\cite{CMS-HIG-14-032}, by the ATLAS and CMS Collaborations, respectively. The \hstarww~channel provides an opportunity for an independent measurement of off-shell Higgs boson production and the Higgs boson width, and it offers a larger sample of events than \hstarzz\xspace, reducing statistical limitations. Studying this channel is also particularly relevant as BSM scenarios may affect \hww~and \hzz~differently. In particular, models that violate custodial symmetry, including scalar extensions of the Standard Model (e.g., multi-Higgs doublet models \cite{solberg2013terms}), can introduce deviations between these channels, making it essential to analyze both in order to fully probe new physics effects \cite{farina2012dys, butterworth2024custodial}. However, the \hstarww~channel also carries additional challenges: the fully-leptonic final state of the \hstarww~channel has substantially higher backgrounds than the \hstarzzllll~channel, and selections on the mass of the vector bosons are less effective than in both the \hstarzzllvv~and \hstarzzllll~channels due to the presence of neutrinos in both $W$ decays.

This paper presents the first LHC Run 2 measurement of off-shell Higgs boson production in the fully leptonic final state of the \hstarww~channel. It also provides the first standalone \hstarww~interpretation of $\Gamma_H$ by the ATLAS experiment. The analysis considers Higgs boson production via gluon--gluon fusion (ggF), \ggfsigeq, and electroweak (EW), \vbfsigeq, production modes. In the off-shell regime, the interference of these processes with non-resonant backgrounds that share the same initial and final states, \ggfbkgeq~and \vbfbkgeq~respectively, is large. The full process in each production mode, including the Higgs-mediated signal, the background, and the interference terms, is referred to as \ggfsbieq~and \vbfsbieq, respectively. Example Feynman diagrams for these processes are shown in Figure~\ref{fig:intro:feynmandiagrams}. Due to the predominantly destructive nature of the interference in the phase space considered in this analysis, the presence of off-shell Higgs boson production is characterized by a deficit in the number of events relative to the background-only hypothesis.

The analysis considers both different-flavour (DF, \hstarwwdf) and same-flavour (SF, \hstarwwsf) leptonic final states, including cases where the $W$ bosons decay via intermediate $\tau$ leptons, introducing one additional $\tau$-neutrino in the final state. Events are categorized based on the decay mode (DF or SF) and the number of jets, $N_j$ (0, 1, or $\geq$2), leading to six distinct analysis regions, each with a dedicated event selection. A classification neural network (DNN) is trained for each region to discriminate against non-interfering backgrounds, and in each region events are divided into three categories based on the DNN score. To enhance sensitivity to the interference effects, an additional variable, \vthreeone, is reconstructed in each of these categories. Defined as a combination of the dilepton invariant mass and the transverse mass of the dilepton-plus-missing-energy system, \vthreeone~serves as a proxy for the non-reconstructable true (``particle-level'') invariant mass of the $WW$ system, $m_{WW}$. A likelihood fit to this variable across all regions and categories is performed to extract the off-shell Higgs boson production signal strength. Finally, a combined fit of this off-shell analysis with the corresponding on-shell measurement~{\cite{HIGP-2024-07}}, using the same final state and data-set, is used to constrain $\Gamma_H$.

%

%

%
%

%
%

%

%
%

% End of text imported from the .//sections/intro.tex input file

%
\section{ATLAS detector}
\label{sec:detector}
%

%

% The next lines are included from the .//sections/detector_letter.tex input file
The ATLAS experiment~\cite{PERF-2007-01} at the LHC \cite{Evans:2008zzb} is a multipurpose particle detector with a forward--backward symmetric cylindrical geometry and a near \(4\pi\) coverage in
solid angle.\footnote{ATLAS uses a right-handed coordinate system with its origin at the nominal interaction point (IP) in the centre of the detector and the \(z\)-axis along the beam pipe. The \(x\)-axis points from the IP to the centre of the LHC ring, and the \(y\)-axis points upwards. Polar coordinates \((r,\phi)\) are used in the transverse plane, \(\phi\) being the azimuthal angle around the \(z\)-axis. The pseudorapidity is defined in terms of the polar angle \(\theta\) as \(\eta = -\ln \tan(\theta/2)\) and is equal to the rapidity $ y = \frac{1}{2} \ln \left( \frac{E + p_z}{E - p_z} \right) $ in the relativistic limit. Angular distance is measured in units of \(\Delta R \equiv \sqrt{(\Delta \eta)^{2} + (\Delta\phi)^{2}}\).}
It consists of an inner tracking detector (ID) surrounded by a thin superconducting solenoid providing a \qty{2}{\tesla} axial magnetic field, electromagnetic and hadronic calorimeters, and a muon spectrometer. The ID covers the pseudorapidity range \(|\eta| < 2.5\). It consists of silicon pixel, silicon microstrip, and transition radiation tracking detectors. Lead/liquid-argon (LAr) sampling calorimeters provide electromagnetic (EM) energy measurements with high granularity within the region \(|\eta|< 3.2\). A steel/scintillator-tile hadronic calorimeter covers the central pseudorapidity range (\(|\eta| < 1.7\)). The endcap and forward regions are instrumented with LAr calorimeters for EM and hadronic energy measurements up to \(|\eta| = 4.9\). The muon spectrometer surrounds the calorimeters and is based on three large superconducting air-core toroidal magnets with eight coils each. The field integral of the toroids ranges between \num{2.0} and \qty{6.0}{\tesla\metre} across most of the detector. The muon spectrometer includes a system of precision tracking chambers up to \(|\eta| = 2.7\) and fast detectors for triggering up to \(|\eta| = 2.4\). The luminosity is measured mainly by the LUCID--2~\cite{LUCID2} detector, which is located close to the beampipe. A two-level trigger system is used to select events~\cite{TRIG-2016-01}. The first-level trigger is implemented in hardware and uses a subset of the detector information to accept events at a rate close to \qty{100}{\kHz}. This is followed by a software-based trigger that reduces the accepted rate of complete events to \qty{1.25}{\kHz} on average depending on the data-taking conditions. A software suite~\cite{SOFT-2022-02} is used in data simulation, in the reconstruction and analysis of real and simulated data, in detector operations, and in the trigger and data acquisition
systems of the experiment.

% End of text imported from the .//sections/detector_letter.tex input file

%
\section{Data and simulated events}
\label{sec:datamc}
%

% The next lines are included from the .//sections/datamc.tex input file
%

%

%

%

%

%

\subsection{Data samples}
\label{sec:datamc:data}

This analysis uses proton--proton ($pp$) collision events collected at a centre-of-mass energy of 13\,\TeV between the years 2015 and 2018, with either a single-electron or single-muon trigger~\cite{TRIG-2018-01,TRIG-2018-05,TRIG-2019-03}. The transverse momentum, $\pt$, thresholds for both electron and muon triggers were set to 26\,\GeV for the majority of Run 2, except the first year where the thresholds were 24 and 20\,\GeV for electron and muon triggers, respectively. Events are only selected for further analysis if they pass ``good-for-physics'' quality criteria, meaning primarily that the relevant detector components are in good operating condition (see Ref. \cite{DAPR-2018-01} for an exact definition). The total integrated luminosity of the collected dataset is 140~fb$^{-1}$~with an uncertainty of 0.83\%~{\cite{DAPR-2021-01,LUCID2}}.

\subsection{Simulated samples}
\label{sec:datamc:mc}

This section describes the simulation of the $pp$ collisions for all the processes used in the analysis. For all processes containing a Higgs boson, the mass of the Higgs boson was set to 125\,\GeV in the generation, and samples were normalised to cross sections computed for a mass of \qty{125.09}{\GeV}~\cite{HIGG-2014-14}.

The dominant ggF off-shell Higgs boson signal process ($gg \to H^* \to WW$, \ggFS), the interfering gluon-induced background process ($gg \to WW$, \ggFB), and the full process including interference ($gg \to (H^* \to) WW$, \ggFSBI) were each generated at leading order (LO) in QCD with up to one additional parton in the final state using \textsc{Sherpa}~2.2.2~{\cite{Bothmann:2019yzt}} and \OPENLOOPS~\cite{Cascioli:2011va, Denner:2016kdg, Buccioni:2017yxi, Buccioni:2019sur}. The \ggFS, \ggFB~and \ggFSBI~processes were separately corrected to next-to-leading order (NLO) in QCD as a function of the generator-level $m_{WW}$ according to the following steps. First, NLO/LO corrections were derived at fixed order in QCD as a function of $m_{WW}$ using \textsc{PowhegBox}~v2~\cite{Nason:2004rx, Frixione:2007vw, Alioli:2010xd} as in Ref.~\cite{Alioli:2021}. The corrections generally decrease as a function of $m_{WW}$, amounting to around 100\% at 200\,\GeV, and flattening out to around 50\% from 400\,\GeV onwards. Second, these corrections were applied to LO MC samples generated using \textsc{MadGraph5}~{\cite{Alwall:2014hca}} and showered using \PYTHIA[8.2]~{\cite{Sjostrand:2014zea}}. Lastly, the $m_{WW}$ spectrum of the \textsc{Sherpa} samples was reweighted to match the corrected \textsc{MadGraph5} spectrum. This method is necessary because the fixed-order NLO/LO corrections derived with \textsc{PowhegBox}~v2 do not apply fully to the merged \textsc{Sherpa} samples. The stepwise approach preserves the per-event kinematic information of the \textsc{Sherpa} samples, while correcting its $m_{WW}$ spectrum differentially to NLO. Two further constant higher-order QCD corrections were applied to the above ggF samples: a factor 1.2 next-to-NLO (NNLO)/NLO correction~\cite{Anastasiou:2005qj, Grazzini:2017mhc, Catani:2007vq}, and a factor 1.1 next-to-NNLO (N$^3$LO)/NNLO correction~\cite{Anastasiou:2015ema}. The former was derived for the Higgs-mediated process, while the latter was derived for a region of phase space where the on-shell component dominates. Corrections that apply more closely to the phase space or samples considered here are not available, so these corrections were applied uniformly for \ggFS, \ggFB~and \ggFSBI.

Following ggF, the next largest source of off-shell Higgs boson production is through EW processes. These include vector boson fusion (VBF) and the associated production of a vector boson and an off-shell Higgs boson (\vhstar), with vector boson decays to quark anti-quark pairs. Four samples were generated to model the EW contribution: the full process including both the signal and its interfering background ($qq \to (H^* \to) WWjj$, EW SBI), a variant of this process with \muoff~and the Higgs boson width both scaled by a factor of 20 relative to the SM (EW SBI20), the interfering background alone ($qq \to WWjj$, EW B), and the signal process excluding the contribution from the $t$-channel Higgs boson diagram ($qq \to H^*\to WWjj$, EW S). Generating a complete signal-only sample is not possible because some diagrams contributing to the EW process lack an $s$-channel Higgs exchange, making it impossible to apply the usual high transverse mass requirement to ensure the Higgs boson is off-shell. Therefore, the EW S sample was used solely for training the neural network as described in Section~\ref{sec:selection}. Combinations of the EW SBI20, EW SBI and EW B samples allow the signal, background and interference components to be accounted for in the full parameterization of the event yield, which is further detailed in Section~\ref{sec:stats}.
For each of these samples, the hard scattering process was generated by \textsc{MadGraph}5~at LO in QCD and EW and the \NNPDF[3.0nlo]~\cite{Ball:2014uwa} set of parton distribution functions (PDFs) was used. The parton shower, hadronisation and underlying event were modelled using \PYTHIA[8.2] with the dipole-recoil scheme and A14 set of tuned parameters (tune)~\cite{ATL-PHYS-PUB-2014-021}. NLO corrections for the EW contributions exist \cite{Denner_2022}, but they are calculated for a different phase space to the one used in this analysis, and are therefore not used.

On-shell Higgs boson production in the \hwwstar~channel was modelled for ggF, VBF, $VH$ and \ttH. The small backgrounds due to on-shell \htautau~decays were also modelled for the ggF, VBF and $VH$ production modes. In both cases, production via ggF was modelled at NNLO in QCD using \textsc{Powheg NNLOPS}~{\cite{Hartanto:2015uka}} and normalised to N$^3$LO in QCD and NLO in EW~\cite{deFlorian:2016spz}. Both VBF and $VH$ modes were generated with \textsc{Powheg Box v2} at NLO in QCD and normalised to NNLO QCD and NLO EW~\cite{Nason:2009ai,Luisoni:2013cuh,deFlorian:2016spz}. The \ttH~mode was modelled using the \POWHEGBOX[v2]~\cite{Frixione:2007nw,Hartanto:2015uka} generator at NLO and normalised to a cross section calculated at NLO in QCD and NLO in EW accuracy~\cite{deFlorian:2016spz}.
The on-shell production modes were interfaced to \PYTHIA[8.212] (for ggF and $VH$) or \PYTHIA[8.230] (for VBF and \ttH) for parton showering and hadronisation, with parameters set according to the AZNLO~\cite{STDM-2012-23} (for ggF, VBF, and $VH$) or A14 (for \ttH) tunes. The samples use the \PDFforLHC[15nnlo]~\cite{Butterworth:2015oua} (for ggF, VBF, and $VH$) or \NNPDF[3.0nlo] (for \ttH) PDF sets.

Fully-leptonic diboson ($WW$, $ZZ$, $WZ$, and $V\gamma^*$) and $V\gamma$ processes were generated with \textsc{Sherpa} 2.2.12 and 2.2.8 respectively, while semi-leptonic final states for $ZZ$ and $WZ$ were generated with \textsc{Sherpa} 2.2.11. The samples are accurate at NLO in QCD for up to one jet and LO in QCD for up to three jets. Loop induced $gg\to ZZ$ events including the off-shell Higgs boson signal, background and interference were generated at LO in QCD with up to one additional parton emission using \textsc{Sherpa}~2.2.2, for fully-leptonic and semi-leptonic final states.  Triboson ($VVV$) processes were modelled by \SHERPA[2.2.2] using factorised gauge-boson decays, and are accurate at NLO in QCD inclusively and LO for up two additional jets.

Events containing a vector boson and jets ($V$+jets) were simulated using \textsc{Sherpa}~2.2.11 (for $ee$ and $\mu\mu$ final states) and \textsc{Sherpa}~2.2.14 (for $\tau\tau$ final states). The simulations are accurate at NLO in QCD for up to two jets and LO in QCD for up to five jets and are calculated in the five-flavour scheme using \textsc{Comix}~\cite{Gleisberg:2008fv}. The samples were normalised to cross sections computed at NNLO in QCD~\cite{Anastasiou:2003ds}.

For all \SHERPA samples, the matrix element calculations were matched and merged with the \SHERPA parton shower based on Catani-Seymour dipole factorisation~\cite{Schumann:2007mg, Gleisberg:2008fv} using the \MEPSatNLO prescription~\cite{Catani:2001cc, Hoeche:2009rj, Hoeche:2011fd, Hoeche:2012yf} and the default tuning parameters. Virtual QCD NLO corrections were provided by the \OPENLOOPS library. The \NNPDF[3.0nnlo] set of PDFs was used, along with the dedicated set of tuned parton shower parameters developed by the \SHERPA authors. For $WW$ and $V$+jets samples, approximate NLO EW corrections were applied following the additive approach~\cite{Kallweit:2014xda}.

Electroweak $Z$ production in association with two jets (EW $Zjj$) was generated using \HERWIG[7.1.3] and \HERWIG[7.2.0]~\cite{Bahr:2008pv, Bellm:2015jjp} for $Z\to{ee}/\tau\tau$ and $Z\to\mu\mu$, respectively. The simulation is accurate at NLO in QCD for up to two additional parton emissions and uses the \MMHT[nlo]~\cite{Harland-Lang:2014zoa} set of PDFs.

Top quark pair production ($t\bar t$) was simulated at NLO in QCD using \textsc{PowhegBox}~v2, and the $h_\text{damp}$ parameter\footnote{The
\hdamp parameter is a resummation damping factor and one of the
parameters that controls the matching of \POWHEG matrix elements to
the parton shower and thus effectively regulates the
high-\pT radiation against which the \ttbar system recoils.}
set to 1.5 times the mass of the top quark~\cite{ATL-PHYS-PUB-2016-020}. The events were interfaced to \PYTHIA[8.230] with the A14 set of tuning parameters and using the \NNPDF[2.3lo] set of PDFs~\cite{Ball:2012cx}. The shape of the $t\bar t$ background was corrected using a recursive generator-level reweighting to NNLO QCD plus NLO electroweak distributions, as described in Ref.~\cite{Czakon:2017wor}. This accounts for known mismodellings of some kinematic variables in the \textsc{PowhegBox}~v2 $t\bar t$ simulation used here. The events were normalised to a cross section computed at NNLO in QCD with NNLL corrections~\cite{Beneke:2011mq, Cacciari:2011hy, Czakon:2011xx, Baernreuther:2012ws, Czakon:2012zr, Czakon:2012pz, Czakon:2013goa}.

Single top quark production in the $s$-channel and associated production of a top quark and $W$~boson ($Wt$) were generated at NLO in QCD using \textsc{PowhegBox}~v2 in the five-flavour scheme and using the \NNPDF[3.0nlo] set of PDFs~\cite{Re:2010bp}. Single top quark production in the $t$-channel was generated at NLO in QCD using \textsc{PowhegBox}~v2 in the four-flavour scheme. The diagram removal scheme~\cite{Frixione:2008yi} was used to remove diagrams that are modelled in both the $tW$ and the $t\bar t$ MC samples. The events were normalised to a cross section computed at NLO in QCD with NNLL soft-gluon corrections~\cite{Kidonakis:2010ux, Kidonakis:2013zqa}.

The response of the detector for the above MC samples was modelled using a full simulation of the ATLAS detector~{\cite{SOFT-2010-01}} based on \textsc{Geant~4}~{\cite{Agostinelli:2002hh}}. \Pileup effects from the same and nearby bunch-crossings were modelled by overlaying minimum-bias hard-scatter events simulated using the soft QCD processes of \PYTHIA[8.1]~{\cite{Sjostrand:2007gs}} with the A3 set of tuned parameters~{\cite{ATL-PHYS-PUB-2016-017}}.
%

% End of text imported from the .//sections/datamc.tex input file

%
\section{Object and event reconstruction}
\label{sec:reco}
%

% The next lines are included from the .//sections/reco.tex input file
This analysis relies on selecting regions with enhanced purity in ggF and EW off-shell \mbox{\hstarww}~production based on the successful reconstruction of collision vertices, electrons, muons, jets, and the missing transverse momentum. Additionally, identification of jets containing $b$-hadrons ($b$-jets) is crucial to suppress top-quark-induced backgrounds. The definitions of these objects in the off-shell analysis are aligned with those of the on-shell \hwwstar~analysis~\cite{HIGP-2024-07}.

Vertices are reconstructed from tracks in the ID with transverse momentum $\pT > 500$\,\MeV, and events are required to have at least one vertex with at least two associated tracks. The primary vertex is selected as the vertex with the highest $\sum \pTsquared$, where the sum is over all the tracks associated with that vertex.

Electron candidates are reconstructed by matching energy clusters in the EM calorimeter to well-reconstructed tracks that are extrapolated to the calorimeter~\cite{EGAM-2018-01}. Electron candidates are required to satisfy $|\eta| < 2.47$, excluding the transition region $1.37 < |\eta| < 1.52$ between the barrel and endcaps of the LAr calorimeter. Muon candidates are reconstructed from a global fit matching tracks from the inner detector and muon spectrometer~\cite{MUON-2018-03}. They are required to satisfy $|\eta| < 2.5$. For both types of leptons, several additional requirements are made to reject particles misidentified as prompt leptons, including likelihood- and selection based identification criteria for electrons and muons, respectively. Electron candidates in events with less than two jets must satisfy the Tight likelihood working point criteria (70\% efficiency), and those in events with two jets or more must satisfy the Medium likelihood working point criteria (85\% efficiency)~\cite{EGAM-2021-01}. Muon candidates must satisfy the Tight identification working point criteria (95\% efficiency)~\cite{MUON-2018-03}. The impact parameter requirements are $|z_0 \sin{\theta}| < 0.5$ mm and $|d_0| / \sigma_{d_0} < 5~(3)$ for electrons (muons).\footnote{The transverse impact parameter, $d_0$, is defined by the point of closest approach of the track to the beamline in the $r-\phi$ plane. The longitudinal impact parameter, $z_0$, is the longitudinal distance to the hard-scatter primary vertex from this point. The uncertainty in the measurement of the transverse impact parameter is denoted by $\sigma_{d_0}$.} Lepton signatures are further required to be isolated from additional energy deposits and tracks in the event. The isolation requirement, defined using a multivariate method based on a boosted decision tree classifier, corresponds to the Tight working point of an improved algorithm relative to that used in Ref.~\cite{HIGG-2017-02}. This updated approach combines isolation variables with lifetime information derived from tracks in a jet reconstructed from the lepton's inner detector tracks, helping to distinguish prompt leptons from those originating in heavy-flavor decays. It achieves more than a factor of two improvement in rejecting misidentified leptons, while maintaining high efficiency for prompt leptons.
Jets are reconstructed using the anti-$k_{t}$ algorithm with a radius parameter of $R$ = 0.4 and particle flow objects as inputs~\cite{Cacciari:2008gp, Fastjet,PERF-2015-09}. Their four-momentum is corrected as explained in Ref.~\cite{PERF-2016-04}. A jet-vertex-tagger multivariate discriminant selection that reduces contamination from \pileup~\cite{PERF-2016-04} is applied to jets with $20 < \pT < 60$\,\GeV and $|\eta| < 2.4$; for jets with $|\eta| > 2.5$, a dedicated forward discriminant~\cite{PERF-2016-06,ATL-PHYS-PUB-2019-026} is used. Jets are required to have $\pT > 20$\,\GeV and $|\eta| < 4.5$, but for the separation of events into jet multiplicity regions only jets with $\pT > 30$\,\GeV are counted. $b$-jets with $\pT > 20$\,\GeV and $|\eta| < 2.5$ are identified using a neural-network discriminant called DL1r~\cite{FTAG-2018-01}. This lower $\pT$ requirement on the $b$-jets compared to jets is adopted to facilitate the definition of a 0-jet top control region. The working point that is adopted has an average 85\% $b$-jet tagging efficiency for simulated $t\bar{t}$ events. At the event selection stage, events are removed if they do not pass the jet cleaning criteria, which are designed to eliminate jets reconstructed from non-collision background processes~\cite{ATLAS-CONF-2015-029}. Finally, standard kinematic and topological criteria are applied to the reconstructed leptons and jets to resolve overlaps, following the procedure outlined in Ref.~\cite{HIGP-2024-07}.

The missing transverse momentum \ptmissvec, with magnitude \ptmiss, is calculated as the negative vector sum of the \pt of all the selected leptons and jets, together with reconstructed tracks that are not associated with these objects but are consistent with originating from the primary $pp$ collision vertex and are referred to as the track soft term ~\cite{JETM-2020-03}. Missing transverse momentum may also arise from the mismeasurement of the momentum of leptons or jets. The \ptmiss significance, \ptmisssig, is used to mitigate this effect, and is calculated from the resolution of the physics objects used in the \ptmiss reconstruction~\cite{JETM-2020-03}.

% End of text imported from the .//sections/reco.tex input file

%
\section{Event selection}
\label{sec:selection}
%

% The next lines are included from the .//sections/selection.tex input file
%
%
%
%
%
%
%

%

A general preselection is first applied to all events for them to be considered in this analysis.
Events are required to contain exactly two oppositely-charged leptons with $\pt > $ 15\,\GeV.\footnote{For the purpose of vetoing additional leptons at the preselection level, the identification and isolation criteria described in Section~\ref{sec:reco} are relaxed to the Medium and FixedCutLoose working points, respectively, for both muons and electrons.} The leading lepton in the event is required to have $p_\text{T}>27\,\GeV$ and to be matched to the object that triggered the event. In the SF channel, events are also required to satisfy $p_{\text{T}}^{\ell\ell}>40\,\GeV$ and $\ptmisssig$ $>4$, with $p_{\text{T}}^{\ell\ell}$ defined as the magnitude of the transverse momentum of the di-lepton system.

The SF and DF channels are further split by jet multiplicity, separating events with 0, 1 and $\geq$ 2 jets. Additional event selections are applied in each of these jet multiplicity regions to define signal regions (SRs) and control regions (CRs), which are further detailed in the following sections.

\subsection{Signal regions}
\label{subsec:signal_reg}

The selections in the three DF SRs vary depending on the jet multiplicity. The invariant mass of the di-lepton system, $m_{\ell\ell}$, is required to be above 100~GeV, 80~GeV, and 70~GeV in the 0-jet, 1-jet, and $\geq$ 2-jets DF SRs, respectively. In the 0-jet DF SR, events with $55 \leq m_{\ell\ell}<100\,\GeV$ are still selected if they satisfy $\Delta\phi_{(\ell,\ell)}>2.0$. The $m_{\ell\ell}$ requirements both here and in the SF SRs ensure orthogonality with the on-shell \hwwstar~analysis, which selects events with $m_{\ell\ell} < 55$ GeV in the 0 and 1-jet multiplicity regions, and $m_{\ell\ell} < 70$ GeV in the 2-jet multiplicity region~\cite{HIGP-2024-07}. Moreover, events in the 0-jet DF SR are required to satisfy $\Delta\eta_{(\ell,\ell)}<1.8$, which cuts out a large proportion of $qq\to WW$ background events. Events in the 1-jet DF SR are required to satisfy $\Delta\phi_{(\ell,\ell)}>1.8$, which facilitates the definition of an orthogonal 1-jet $qq\to WW$ CR while retaining most off-shell Higgs boson events in the SR. Events in all $\geq$ 2-jet SRs and CRs are required to satisfy the central jet veto (CJV) and outside lepton veto (OLV) requirements, which capitalize on the characteristic forward jets of the VBF process. The CJV ensures that the event does not contain a third jet within the pseudorapidity gap spanned by the two leading jets, reducing backgrounds from processes with QCD radiation, whereas the OLV requires that the leading leptons fall within this gap. Lastly, events in all SRs are required not to contain $b$-jets.

In the 0-jet and 1-jet SF SRs, $m_{\ell\ell}$ is required to be above 55~GeV, and in the 2-jet SF SR it is required to be above 70~GeV. In the 0-jet and 1-jet SF SRs, a selection of $\Delta\phi_{(\ell,\ell)}>1.8$ is made, and this requirement ensures orthogonality with the on-shell \hwwstar~analysis~\cite{HIGP-2024-07}. Events in all SF SRs are required to contain no $b$-jets, and must satisfy $\vert m_{\ell\ell}-91\,\GeV\vert>15\,\GeV$ to reject background processes containing $Z$~bosons. The CJV and OLV are applied in the 2-jet SF SR. For a potential future combination, the events in all three SF SRs must satisfy $\Delta R_{(\ell,\ell)}>1.8$. This selection ensures orthogonality with the $ZZ \to \ell \ell \nu \nu$ channel of the off-shell \hstarzz~analysis~\cite{HIGP-2024-14}, where the two leptons typically have a smaller angular separation as they originate from the same vector boson, in contrast to the \hstarww~channel.

\subsection{Signal-background separation}

DNN-based classifiers built with Tensorflow~\cite{tensorflow2015-whitepaper} are trained separately in each SR and are used to divide events into three DNN-score categories per SR, where the division was chosen to optimise expected analysis sensitivity. These DNNs are trained to separate both the full process (SBI) and the signal (S) from the non-interfering backgrounds listed in Section \ref{sec:background}. For each SR, a subset of the following variables is provided as feature variables to the DNN: $p_\text{T}$, $\eta$, $\phi$, and $m$ of the leading and subleading lepton ($\ell_0$ and $\ell_1$)\footnote{The lepton masses are not included.} and jets ($j_0$ and $j_1$); $p_{\text{T}}^{\ell\ell}$; $|\Delta\eta_{(\ell,\ell)}|$; $|\Delta y_{(\ell,\ell)}|$; $|\Delta\phi_{(\ell,\ell)}|$; $\Delta R_{(\ell,\ell)}$; $m_{\ell\ell}$; $m_\text{T}$, the transverse mass, which is defined as $m_\text{T} = \sqrt{(E_{\text{T}}^{\ell\ell}+\ptmiss)^2-\vert\vec{p}_{\text{T}}^{\ell \ell} + \ptmissvec \vert^2}$, where $E_{\ell\ell}=\sqrt{p_{\text{T}}^{\ell\ell\,2}+m_{\ell\ell}^2}$; $m_{jj}$, the invariant mass of the two leading jets; $|\Delta y_{(j,j)}|$, computed from the two leading jets; $\sqrt{H_\text{T}}$, where $H_\text{T}$ is the sum of the scalar transverse momenta of preselected leptons and jets in the event; $\phi_{\ptmissvec}$; \ptmisssig; and $\Delta R_{(i,k)}$, where the $ik$ combinations used are $\ell_0j_0$, $\ell_0j_1$ and $\ell_1j_0$. In the 0-jet SR, a combination of variables associated to the leptons, $E_\text{T}^\text{miss}$ and event variables were chosen as DNN input variables. In the 1-jet and 2-jet SRs, permutation tests are used to select highly performing input variables. All of the MC predictions of DNNs and their input variables are consistent with data within the statistical and systematic uncertainties across all CRs described in Section \ref{subsec:control_regions}.

In the training of the DNN, a simulated sample of $W$+jets events is used to model the contribution from misidentified lepton backgrounds. The training targets are ggF S and ggF SBI in the 0-jet regions, EW S and EW SBI in the 2-jet regions, and all of these contributions combined in the 1-jet regions. For both ggF and EW, the SBI and S processes are normalized equally in the SRs, ensuring that the model focuses also on S, which would otherwise contribute only minimally due to its relatively small yield. The combined sum of weights of all targets are scaled to equal the total sum of background weights in the training. This reweighting balances the otherwise very imbalanced classes, which aids in the training. Hyperparameter optimisations are performed separately for each DNN to maximise a significance-based sensitivity metric. A five-fold cross-validation\footnote{For each task, five independent DNNs are trained using a five-fold cross-validation strategy. The simulated events are split into five equal subsets; each DNN is trained on three, validated on one, and tested on the remaining subset. The roles of the subsets are rotated to ensure statistical independence between training, validation, and testing. Only the inference on the test subset, unused during training or validation, is used in the analysis.} is used in the training and inference~\cite{kohavi1995study}.

Due to the presence of destructive interference, the contributions to the event yields from the signal and interference may partially or completely cancel out, depending on the specific region of the phase space. However, they both contribute to the analysis sensitivity. As such, it is paramount to find a discriminant achieving high separation between signal and interference. The particle-level invariant mass of the $WW$ system, $m_{WW}$, provides such separation to some extent, but due to the presence of the neutrinos in the final state, it is impossible to fully reconstruct $m_{WW}$. Instead, this analysis uses a linear combination of $m_{\ell \ell}$ and $m_\text{T}$, $\vthreeone = 0.3 \cdot m_{\ell \ell} + 1 \cdot m_\text{T}$, which provides a simple proxy for $m_{WW}$. The coefficient 0.3 and 1 were determined to be optimal in their $m_{WW}$-regression effectiveness by scanning for both a range of values between 0 and 1.

\subsection{Control regions}
\label{subsec:control_regions}
The CRs are defined to estimate the normalisations of the leading background processes, and reduce the effect of systematic uncertainties by causing a partial cancellation of their impacts in the SRs. The selections that define all CRs are applied in addition to the DF pre-selection requirements; 1-jet CRs are defined for the $qq\to WW$ and $Z\to\tau\tau$ processes, and 0, 1 and $\geq$ 2-jet CRs are defined for the top quark backgrounds. For the $Z\to\tau\tau$ and $qq\to WW$ processes in SRs characterized by other jet multiplicities, and for the $Z\to\ell\ell$ process in all SRs, no corresponding CR is defined as the background-rich low DNN-score categories in the SRs provide sufficient constraining power.

The $qq\to WW$ CR is required to contain no $b$-jets and to have $m_{\ell\ell}$ above 80~GeV. Events are also required to satisfy $\Delta\phi_{(\ell,\ell)}<1.8$ as the $qq \to WW$ background exhibits a more uniform distribution compared to the ggF S process as a function of $\Delta\phi_{(\ell,\ell)}$, and $\vert m_{\tau\tau}-91\,\GeV\vert>25\,\GeV$, where $m_{\tau\tau}$ represents the assumed mass of the $\tau\tau$ system computed from the light leptons and \ptmiss~under the collinear approximation.\footnote{The collinear approximation for $m_{\tau\tau}$ assumes the neutrinos travel in the same direction as the visible $\tau$ decay products, as detailed in Ref.~\cite{Plehn:2000}.} Events for which this approximation yields no physical solution are also vetoed. The purity of $qq\to WW$ events in this CR is estimated to be 28\% using simulations.

The top quark CRs constrain the normalisation of the $t\bar t$ and $Wt$ backgrounds. Events in the 0-jet and 1-jet ($\geq$2-jet) CRs are required to have $m_{\ell\ell}$ above 100~GeV (70~GeV). In the 0-jet CR, there must be one $b$-jet with $20<p_\text{T}<30\,\GeV$, and in the 1-jet and 2-jet CRs there must be exactly one $b$-jet with $p_\text{T}>30\,\GeV$. Events in the 0-jet CR are required to satisfy $\Delta\eta_{(\ell,\ell)}<1.8$, and events in the 1-jet CR are required to satisfy $\Delta\phi_{(\ell,\ell)}>1.8$. The 2-jet CR requires events to satisfy the CJV and OLV. The purity of top quark events is 93\% in the 0-jet top quark CR and 98\% in both the 1- and 2-jet top quark CRs as estimated using simulations.

Events in the $Z\to\tau\tau$ CR are required to satisfy $m_{\ell\ell}<80\,\GeV$ and not to contain any $b$-jets. Events in this CR are also required to have $m_{\tau\tau}$ larger than $66\,\GeV$. The purity of $Z\to\tau\tau$ events in this CR is estimated to be 88\% using simulations.

%

%

%
%
%
%
%
%
%
%
%
%
%
%
%
%
%
%
%
%
%
%
%
%
%
%
%
%
%
%
%

%
%
%
%
%
%

%
%
%
%
%
%
%
%

%
%
%
%
%
%
%

%
%
%
%
%
%
%

%
%
%
%
%
%
%
%
%
%

%
%
%
%
%
%
%
%
%
%
%
%

%
%
%
%
%
%
%
%
%
%
%

%

%

% End of text imported from the .//sections/selection.tex input file

%
\section{Background estimation}
\label{sec:background}
%

% The next lines are included from the .//sections/background.tex input file
%

%

This analysis is subject to two types of backgrounds: those that interfere with the signal and those that do not. The former were introduced in Section~\ref{sec:intro}, while the latter include events from $qq\to WW$ processes, top-quark-induced backgrounds ($t\bar{t}$ and $Wt$), $Z\to\ell\ell$ and $Z\to\tau\tau$ decays, $VVV$ events, $WZ$ events, $V\gamma^*$ events, a tiny contribution from on-shell Higgs boson processes including $WH\to \ell\nu qq \ell\nu$,  and events involving misidentified leptons. The normalisations of the $qq\to WW$, top-quark-induced, $Z\to\tau\tau$, and $Z\to\ell\ell$ backgrounds are estimated and constrained using data from the relevant CRs, described in Section~\ref{sec:selection}, or low DNN-score categories in the SRs. Other background normalisations are derived from simulation except for the misidentified lepton background, which requires a dedicated data-driven approach.

The misidentified lepton background arises mainly from $W$+jets events, where jets are incorrectly identified as prompt leptons. The probability of such misidentification is small, but the large cross section of the $W$+jets process compared to that of the signal means that this background must still be taken into account. The normalisation and shape of this background are estimated using the data-driven fake-factor method~\cite{LEHMANN2023168376}, since it is not well modelled in simulation. This method uses an orthogonal misidentified-lepton-enriched control region, in which all nominal selections are applied except that at least one lepton candidate fails nominal lepton identification and isolation criteria but instead passes a looser set of criteria. The measured misidentified lepton background is estimated by subtracting the contribution from prompt background processes from data in the misidentified-lepton-enriched control region, and extrapolating to the SR using fake factors. These fake factors are evaluated as a function of \pt and $|\eta|$ in a region dominated by $Z$+jets events. The derivation of the fake factors is described in more detail in Ref.~\cite{HIGP-2024-07}, in the context of the on-shell analysis.

%
%

%

% End of text imported from the .//sections/background.tex input file

%
\section{Systematic uncertainties}
\label{sec:systematics}
%

% The next lines are included from the .//sections/systematics.tex input file
The sensitivity of the analysis is impacted by several sources of systematic uncertainty that are related to the theoretical description of the simulated samples and the experimental uncertainties originating from the detector response.

Each analysis object defined in Section~\ref{sec:reco} has associated experimental uncertainties that are applied to the objects in MC simulations. For leptons, the uncertainties are related to the trigger, reconstruction, identification, and isolation selection efficiencies as well as the energy (or momentum) scale and resolution. For jets, uncertainties arise from the jet energy scale and resolution, the jet-vertex tagger performance, and the $b$-jet identification. Another source of uncertainty originates from the impact of the computation of the track soft term in the reconstruction of \ptmiss. The uncertainty in the modelling of pile-up for simulated samples is estimated by varying the reweighting to the profile in data within its uncertainties. The uncertainty in the combined integrated luminosity is 0.83\%~\cite{DAPR-2021-01}, obtained using the LUCID-2 detector~\cite{LUCID2} for the primary luminosity measurements.

The uncertainties in the misidentified lepton estimate, described in Section~\ref{sec:background}, have three main components: the statistical uncertainty in extrapolation factors derived in dedicated misidentified-lepton-enriched regions, the uncertainty from subtracting the simulated prompt lepton component, and the sample composition uncertainty detailed in Ref.~\cite{HIGP-2024-07}. The relative size of the uncertainty on the fake extrapolation factor depends on the lepton flavour and its momentum, and ranges roughly between 10\% and 70\%.
For all signal and major background processes, theory uncertainties related to the variation of the nominal choice of PDF, the QCD renormalisation and factorisation scales, and the strong coupling constant $\alpha_{s}$ are evaluated. For \ggFB, \ggFSBI, and \ggFS, these uncertainties are derived on the NLO fixed-order prediction used to obtain the k-factors discussed in Section~\ref{sec:datamc:mc} and are correlated across all three samples.

For processes generated with \textsc{Sherpa}, uncertainties are evaluated in the choice of the matching scale between matrix element calculation and parton shower (CKKW \cite{Catani:2001cc}), the modelling of parton shower effects that include variation of the resummation scale (QSF), and the choice of parton shower recoil scheme (CSSKIN \cite{Hoeche:2009rj, Schumann:2007mg}). All three uncertainties are derived using particle-level samples.

Uncertainties in the parton shower simulation of EW SBI, EW SBI20, and EW B are estimated by comparing the expected yield from the corresponding simulated samples with that from alternative samples where the parton  shower is simulated with \textsc{Herwig} instead of \textsc{Pythia}. In addition, these samples include uncertainties in the modelling of the hard emissions and variations of the showering scale controlling the amount of radiation in both the initial and final states. Since these processes are only simulated at LO, an additional conservative constant 25\% uncertainty is applied to cover both electroweak and QCD NLO corrections~\cite{Denner_2022}.\footnote{The k-factors in the paper provided are derived differentially and inclusively in a phase space that is not identical to the measured phase space in this analysis. Therefore, the size of the uncertainty was chosen as the largest correction provided in Ref.~\cite{Denner_2022}.} A constant 20\% uncertainty is used for the remaining small contributions of on-shell ggF and VBF $H \to WW^*$ and a 50\% uncertainty for other minor Higgs boson production and decay modes.

For processes including top quarks, the NLO matching scheme uncertainty is applied and a comparison with the alternative generator \textsc{Herwig} is used to cover differences in the showering and hadronisation schemes.
In addition, uncertainties related to the modelling of initial- and final-state radiation are represented by varying scale, resummation, and showering parameters. The ordering choice in the 3D recursive reweighting is considered as an additional source of uncertainty for the $t\bar t$ process. For the $Wt$ process, uncertainties related to the choice of $h_{\mathrm{damp}}$ scale~\cite{ATL-PHYS-PUB-2021-042} and diagram removal schemes~\cite{Frixione:2008yi} are applied.

The non-interfering single and diboson background \textsc{Sherpa} samples are affected by an uncertainty in the combination of the NLO QCD with NLO EW virtual approximate correction. An additional generator comparison uncertainty is assigned to cover an observed $p_{T}^{\ell \ell}$ shape mismodelling in the \textsc{Sherpa} 2.2.11/14 Z+jets processes~\cite{HIGG-2020-20}. For triboson processes, a constant 12\% theory uncertainty is used.

The systematic uncertainties alter the total number of expected events (normalisation) and shape of the $\vthreeone$ distributions. To remove negligible contributions and improve the stability of the fit, the systematics uncertainties are pruned, smoothed, and symmetrised. For the shape component of the uncertainties, only the theory uncertainties affecting $Z\to \ell \ell$, $Z\to \tau \tau$, ggF S, B, and SBI, as well as EW B, SBI, and SBI20 are retained, while all other shape uncertainties (from other sources or for other samples) are removed.

The systematic uncertainties described in this section are related to the off-shell signal strength measurement. When translating the signal strength in terms of the Higgs boson width, by combining with the on-shell results as described in Section~\ref{sec:result}, several systematic uncertainties (partially) cancel.

% End of text imported from the .//sections/systematics.tex input file

%
\section{Statistical inference}
\label{sec:stats}
%

% The next lines are included from the .//sections/stats.tex input file
%

To quantify the strength of the signal in the statistical model, the parameter of interest, $\mu$, is measured in this analysis using the profile log-likelihood ratio~\cite{Cowan:2010js} as test statistic:
\begin{align*}
t_{\mu} = -2 \mathrm{ln} \frac{\lambda(\mu, \hat{\hat{\theta}}) }{\lambda(\hat{\mu}, \hat{\theta})};\text{ which generalizes to }t_{\mu_1,\mu_2} = -2 \mathrm{ln} \frac{\lambda(\mu_1, \mu_2, \hat{\hat{\theta}}) }{\lambda(\hat{\mu_1}, \hat{\mu_2}, \hat{\theta})}
\end{align*}
for cases with two parameters of interest, $\mu_1$ and $\mu_2$. Here, $\lambda$ denotes the binned likelihood function, defined as the product of Poisson probability terms over all bins of \(\vthreeone\) in the signal regions (SRs) and single-bin control regions (CRs) considered in the analysis, as introduced in Section~\ref{sec:selection}. The likelihood function depends on the parameter(s) of interest and a set of nuisance parameters (NPs), $\theta$, which account for systematic uncertainties discussed in Section~\ref{sec:systematics}. The maximum likelihood estimates of the parameters are denoted by $\hat{\mu}$ (equivalently $\hat{\mu_1}$ and $\hat{\mu_2}$) and $\hat{\theta}$, while $\hat{\hat{\theta}}$ represents the values of the nuisance parameters that maximize the likelihood function for a given $\mu$ (or equivalently $\mu_1$ and $\mu_2$).

The full ggF signal template, referred to as $\nu^{\mathrm{ggF}}(\muoffggf,\theta)$, for a given hypothesis \muoff{} can be computed for each bin of the input distributions using the following parameterisation:
\begin{align*}
\nu^{\mathrm{ggF}}(\muoffggf,\theta)
&= \muoffggf             \cdot      n_{\mathrm{S}}^{\mathrm{ggF}}(\theta)
+ \sqrt{\muoffggf}    \cdot    \left( n_{\mathrm{SBI}}^{\mathrm{ggF}}(\theta)  -  n_{\mathrm{S}}^{\mathrm{ggF}}(\theta)  -  n_{\mathrm{B}}^{\mathrm{ggF}}(\theta)  \right)
+    n_{\mathrm{B}}^{\mathrm{ggF}}(\theta)  \nonumber \\
&=  \left(  \muoffggf   -    \sqrt{\muoffggf}   \right) \cdot   n_{\mathrm{S}}^{\mathrm{ggF}}(\theta)
+  \sqrt{\muoffggf} \cdot n_{\mathrm{SBI}}^{\mathrm{ggF}}(\theta)
+  \left(  1-   \sqrt{\muoffggf}   \right) \cdot  n_{\mathrm{B}}^{\mathrm{ggF}}(\theta), \nonumber
\end{align*}

where $n_{\mathrm{S}}^{\mathrm{ggF}}$, $n_{\mathrm{B}}^{\mathrm{ggF}}$ and $n_{\mathrm{SBI}}^{\mathrm{ggF}}$ represent the corresponding expected yields of the \ggFS, \ggFB, and \ggFSBI{} processes, respectively. The interplay between signal and destructive interference leads to a non-monotonic total event yield as a function of $\muoff$. This can cause double minima in likelihood scans, a degeneracy that is lifted in part or fully by the information contained in the shape of the fitted \vthreeone~distributions. The definition of an off-shell signal-only template for the EW process is complicated by the presence of non-removable on-shell Higgs contributions as discussed in Section~\ref{sec:datamc:mc}. Instead, the EW parameterisation uses $n_{\mathrm{B}}^{\mathrm{EW}}$ and two $n_{\mathrm{SBI}}^{\mathrm{EW}}$ samples with $\mu^{\text{EW}}_{\text{off-shell}}$ scaled to 1 and 20 respectively. A detailed description of this parametrisation can be found in Table~1 of Ref.~\cite{HIGP-2024-14}, where it was used in the $H \to ZZ$ channel.

All experimental uncertainties, including those from misidentified leptons, are correlated across all measured analysis regions and categories. Theory uncertainties in ggF S, B, and SBI
and separately in EW B, SBI, and SBI20 are also correlated across all analysis regions and categories whereas uncertainties in the non-interfering background are only correlated across lepton flavour channels but not across jet multiplicities. The exceptions are $V\gamma$, diboson processes (not including $qq\to WW$), and triboson processes which are also not correlated across lepton flavours due to the differences in the origin of the leptons in the final state across these channels.

The likelihood can be parameterised as a function of a single off-shell signal-strength parameter $\muoff$ that is applied to both Higgs boson production modes (ggF and EW), but alternatively two signal strength parameters for the ggF and EW process can be fully decorrelated and used separately: $\muoffggf$ and $\muoffew$.

The $\muoff$ measurement is combined with the measurement of $\mu_{\mathrm{on-shell}}$ from Ref.~\cite{HIGP-2024-07} in a joint likelihood function. The result can be interpreted in terms of the width of the Higgs boson normalised to its SM expectation $(\muoff / \muonshell = \Gamma_{H}/\Gamma_{H}^{\mathrm{SM}} = \kappa_H)$ or as the ratio of coupling modifiers in the off-shell and on-shell regimes $(R_{gg} = \kappaGoffshell^2 / \kappaGonshell^2)$ and $(R_{VV} = \kappaVoffshell^2/\kappaVonshell^2)$. The $\kappa_H$ interpretation assumes that the off- and on-shell coupling modifiers are the same for both ggF and EW production modes, while the $R_{gg}$ and $R_{VV}$ interpretations assume that the total width of the Higgs boson is equal to its SM prediction.

In the combination, nuisance parameters representing uncertainties from experimental sources and theory predictions in non-interfering backgrounds in the on- and off-shell analyses are correlated. The theory uncertainties in the interfering backgrounds are not correlated across the on- and off-shell phase spaces, as this would imply also correlating them with the uncertainties in the off-shell signal. The small contribution of on-shell signal in the off-shell analysis was correlated with the signal in the on-shell analysis. An extra selection is made in the on-shell $qq\to WW$ CR at $m_{\ell\ell} < 80$\,\GeV in the combination. This requirement removes the overlap between the $qq\to WW$ CRs in the 1-jet phase space of both analyses. Moreover, the $Z\to\tau\tau$ CR fully overlaps with the corresponding CR of the on-shell analysis. In the combination, an extra selection is therefore applied to the off-shell CR which requires $ 60 \leq p_{\text{T}}^H< 120$\,\GeV, where $p_{\text{T}}^H$ is defined as $\left| \vec{p}_{\text{T}}^{\ell \ell} + \ptmissvec \right|$. This fully aligns the CR definitions, creating a single CR that is shared between the two analyses.

%

%

%

%

%
%

%

%

% End of text imported from the .//sections/stats.tex input file

%
%
%
%

%

%
\section{Results}
\label{sec:result}
%

\begin{figure}[!t]
\centering
\includegraphics[width=.95\textwidth]{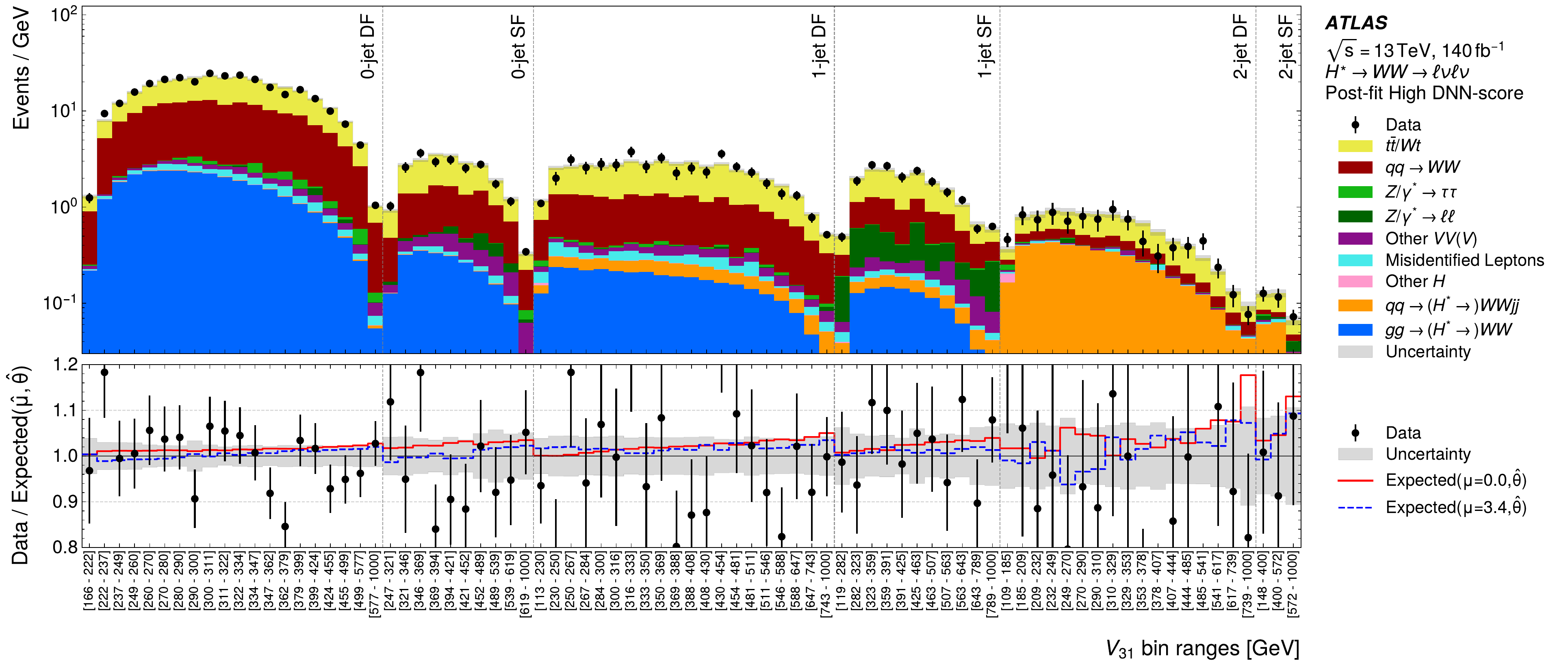} \\
\caption{Summary of the post-fit $V_{31}$ distributions in the high DNN-score category of the SRs. The top panel shows the event yield divided by the bin width in each of the measured categories for both measured data (dots) and expected post-fit predictions (histograms). Overflow is included in the last bin of each distribution. The bottom panel shows the ratio of the data to the post-fit expected predictions. The sum in quadrature of statistical and systematic post-fit uncertainties are depicted by a grey band. Two alternative hypotheses are included, where the nuisance parameters are fixed to their best-fit values, $\hat{\theta}$, and the value of the parameter of interest \muoff = 0 (solid line) and \muoff = 3.4 (dashed line). The former represents the null hypothesis whereas the latter represents the upper bound of the symmetric 95\% CL interval of \muoff.}
\label{fig:dataVSmc_postfit_dnnsummary23}
\end{figure}

% The next lines are included from the .//sections/results.tex input file
Using the likelihood function described in Section~\ref{sec:stats}, a combined fit is performed on all analysis regions and categories. Figure~\ref{fig:dataVSmc_postfit_dnnsummary23} shows the post-fit distributions for the $\vthreeone$ observable in the high DNN-score category of all signal regions.\footnote{The binning optimisation of $\vthreeone$ is mostly driven by the requirement to have more than 1 and 10 signal and non-interfering background events per bin, respectively, and a relative statistical uncertainty below 20\% in each bin. In addition, the width of each bin is required to be above 10\,\GeV, and the maximum number of bins is imposed not to be larger than 20 in the high DNN-score categories to reduce the complexity of the fit. These criteria were varied to study the robustness of the binning and a minimal impact on the results was found.} Two alternate hypotheses are shown in the ratio pad at the bottom, representing the expected signal yield at $\muoff=0$ and $\muoff=3.4$, where it crosses the 95\% upper bound of the symmetric CL interval. In some bins both hypotheses vary in the same direction (up or down), reflecting the nonlinear parametrisation of the signal. The most sensitive regions are the 0-jet multiplicity regions due to the large event yields, followed by the 2-jet multiplicity regions due to their relatively high signal purities.

\begin{figure}[!htbp]
\centering
\begin{subfigure}[b]{0.48\textwidth}
\centering
\includegraphics[width=\textwidth]{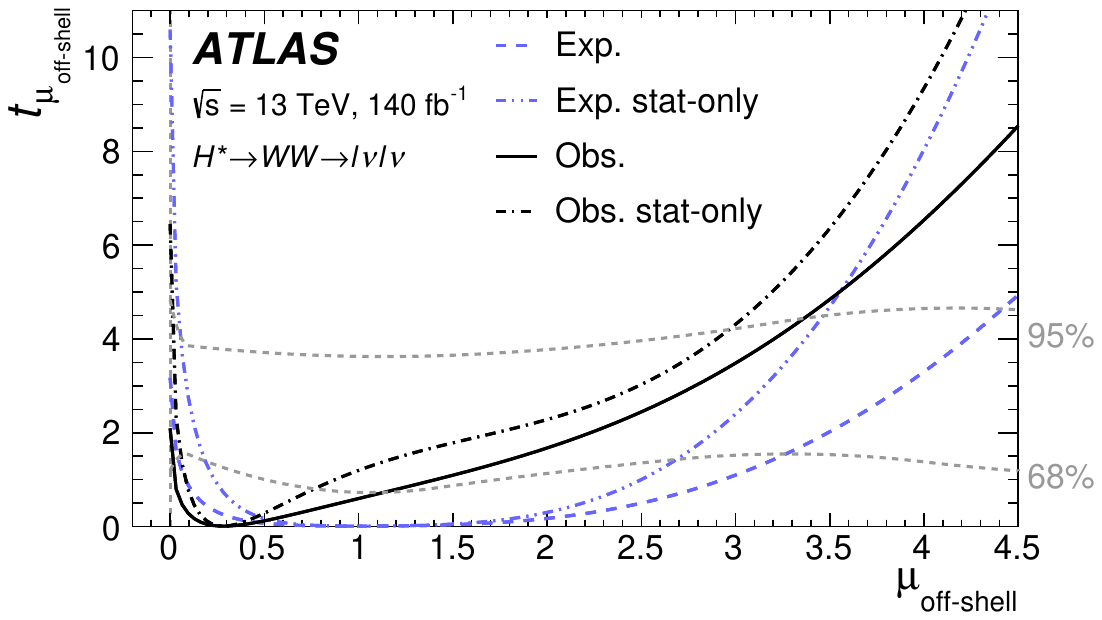}
\caption{}
\label{fig:results_muoffshell:a}
\end{subfigure}
\hfill
\begin{subfigure}[b]{0.48\textwidth}
\centering
\includegraphics[width=\textwidth]{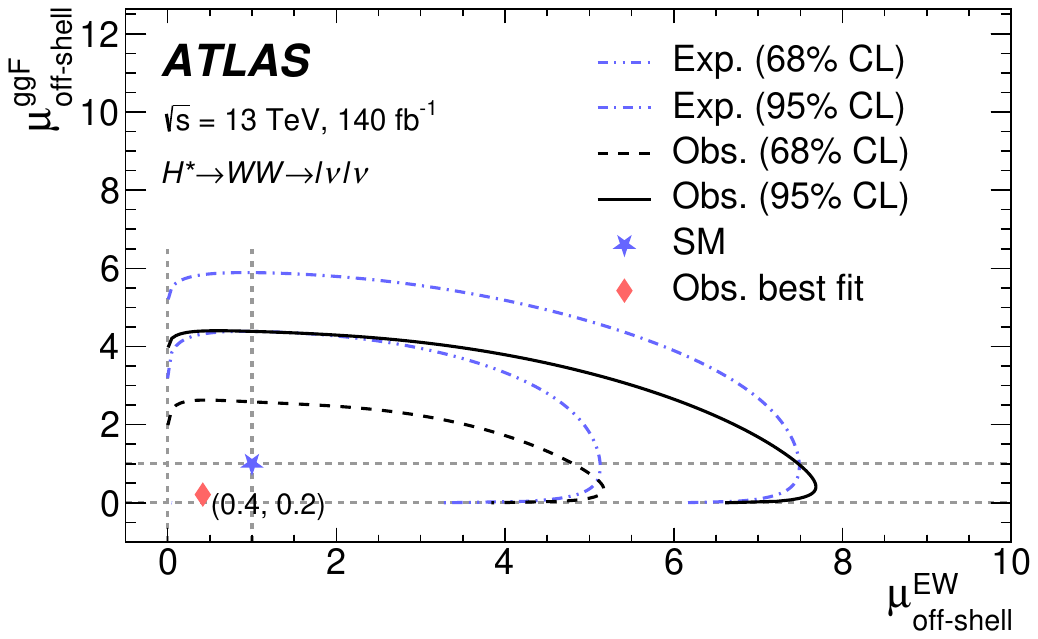}
\caption{}
\label{fig:results_muoffshell:b}
\end{subfigure}

\caption{(a) Values of the test statistic $t_{\mu_{\mathrm{off-shell}}}$ as a function of the off-shell Higgs boson signal strength, $\muoff$. The 68\% and 95\% CL are shown as gray dotted lines. (b) Contours of the values of the profile likelihood ratio $t_{\muoffggf,\muoffew}$ as a function of $\muoffggf$ and $\muoffew$. The SM predicted value and best-fit value are indicated by a star and a diamond, respectively. The observed (expected) results are shown in black (blue) for scenarios with and without systematics. For (a) the 68\% and 95\% CL are calculated using an explicit Neyman construction, whereas in (b) they are calculated via the asymptotic approximation.}
\label{fig:results_muoffshell}
\end{figure}

The values of the profile likelihood ratio as a function of $\muoff$ are shown in Figure~\ref{fig:results_muoffshell:a}, both for the expected and for the observed results. Both of these also include the result with systematic uncertainties fixed to their best-fit values from a statistics-only fit, and the result where they are allowed to vary. The difference between the observed and expected values of the signal strength parameter $\muoff$ can be understood from degeneracies caused by the non-trivial (quadratic) dependence of the event yield as a function on $\muoff$. The quadratic dependence of signal and interference yields on $\muoff$ in the presence of over- or underfluctuations causes the distribution of the test statistic to depart from the asymptotic $\chi^2$ distribution predicted by Wilks' theorem~\cite{wilks1938large}. Therefore, the confidence intervals on $\muoff$ are constructed using the Neyman construction~\cite{neyman1937outline}. These intervals are derived from the distribution of the profile likelihood ratio, evaluated for different values of the parameter of interest. The distribution is obtained through simulated pseudo-experiments, where event yields in each bin are sampled from the probability density function with nuisance parameters fixed to their best-fit values from data. The measurement yields observed (expected) central values and 68\% confidence level (CL) uncertainties on the signal strength of $\mu_{\mathrm{off-shell}} = 0.3^{+0.9}_{-0.3} (1.0^{+2.3}_{-1.0})$. The observed (expected) upper bound on the measurement at 95\% CL on $\muoff$ is 3.4 (4.4). Table~\ref{table:systbreakdown} shows the observed impact of groups of systematic uncertainties for the value of \muoff{} at $t_{\muoff}=1$ following the procedure described in Table~5 of Ref.~\cite{HIGG-2018-32}. The leading uncertainty is driven by the statistical precision of the available dataset, followed by theory and experimental uncertainties. In the \muoff~fit, all estimated background normalisations are compatible with the SM predictions within 95\% CL.

\begin{table}[!t]
\centering
\caption{Breakdown of the observed impact of sources of uncertainties on the value of $\muoff$ at 68$\%$ confidence level where $t_{\muoff}=1$. The values in the right column represent the relative difference in quadrature between the best-fit $\muoff$ and the $\muoff$ from a fit where a set of nuisance parameters ($\theta_i$) are fixed to their best-fit values $\hat{\theta}_{i}$.}
\label{table:systbreakdown}
\begin{tabular}{ l | c  }
\hline
\hline
Statistical uncertainty     & 52\%   \\
MC stat. uncertainty           & 15\%   \\
Theory uncertainty             & 39\%   \\
~~~~~- Theory background            & 22\%   \\
~~~~~- Theory signal                & 34\%   \\
Experimental uncertainty       & 25\%   \\
~~~~~- Jets                         & 19\%   \\
~~~~~- Leptons                      & 5.3\%   \\
~~~~~- Others                       & 6.8\%   \\
~~~~~- Misidentified leptons                   & 3.1\%   \\
Background normalisation       & 7.6\%   \\
\hline
\hline
\end{tabular}
\end{table}

To probe the two off-shell signal strength parameters for the ggF and EW production modes, Figure~\ref{fig:results_muoffshell:b} shows the values of the profile likelihood ratio as a function of both $\muoffggf$ and  $\muoffew$. The best-fit values for the parameters are observed (expected) to be
$\mu^{\mathrm{ggF}}_{\mathrm{off-shell}} = 0.2^{+1.3}_{-0.2} ~(1.0^{+2.5}_{-1.0})$ and
$\mu^{\mathrm{EW}}_{\mathrm{off-shell}} = 0.4^{+3.4}_{-0.4} ~(1.0^{+2.9}_{-1.0})$.

\subsubsection*{Higgs boson width measurement}
\begin{figure}[!htbp]
\centering
\includegraphics[width=0.6\textwidth]{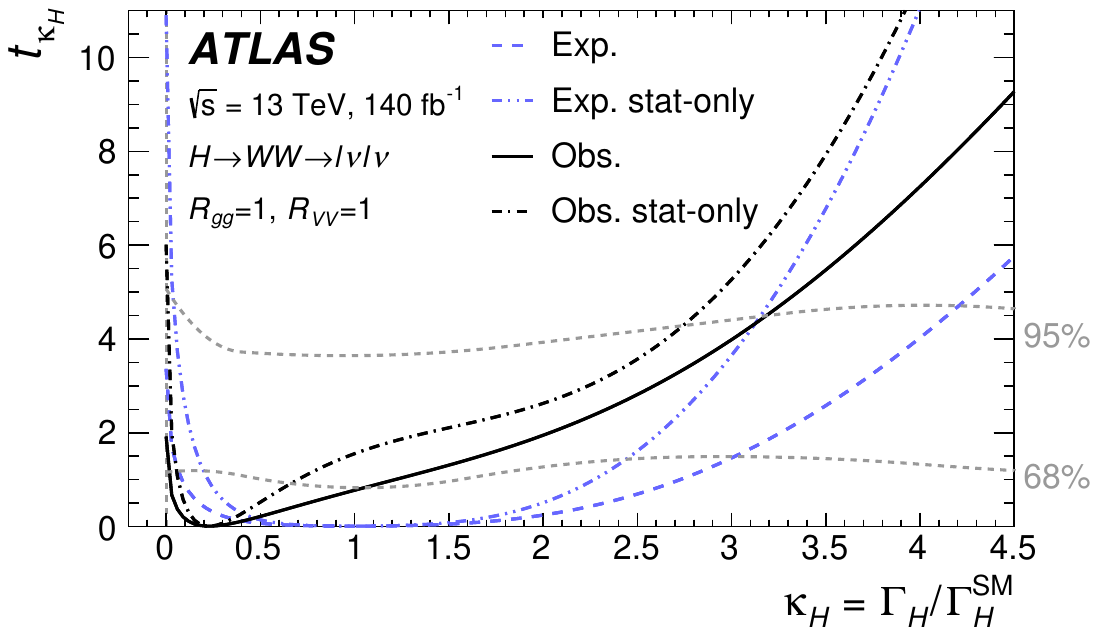}
\caption{Values of the test statistic $t_{\kappa_H}$ as a function of $\kappa_H = \gammaH / \gammaHSM$ for the combined on-shell and off-shell result assuming $R_{gg}=R_{VV}=1$. The observed (expected) results are shown in black (blue) for scenarios with and without systematics. The 68\% and 95\% CL are shown as gray dotted lines. The 68\% and 95\% CL were calculated using an explicit Neyman construction.}
\label{fig:results_width}
\end{figure}

As described in Section~\ref{sec:stats}, the combination of this analysis with the on-shell \hwwstar~analysis~\cite{HIGP-2024-07} allows for an interpretation in terms of the Higgs width normalized to its SM value, $\kappa_H$, or alternatively the ratio of the Higgs to $gg$ and $VV$ couplings squared in the on- and off-shell regimes, $R_{gg}$ and $R_{VV}$. The resulting values of the test statistic $t_{\kappa_H}$ are shown in Figure~\ref{fig:results_width} as a function of $\kappa_H$. Using $\Gamma_H^{\mathrm{SM}}=4.1$~MeV, the observed (expected) value for $\gammaH$ is $0.9^{+3.4}_{-0.9}$ ($4.1^{+8.3}_{-3.8}$~\MeV) and the observed (expected) upper bound at 95\% confidence level on $\gammaH$ is 13.1 (17.3)~\MeV. The interpretation in terms of $R_{VV}$ yields an observed (expected) value of $0.6^{+1.6}_{-0.6}$ ($1.0^{+1.2}_{-1.0}$), while the $R_{gg}$ interpretation is not sufficiently sensitive to set upper bounds at 68\% as the likelihood function rapidly flattens. In all interpretations, the estimated background normalisations are compatible with those from the \muoff~fit within 68\% CL.

\FloatBarrier

% End of text imported from the .//sections/results.tex input file

%
%
%

%
\section{Conclusion}
\label{sec:conclusion}
%

% The next lines are included from the .//sections/conclusions.tex input file
This Letter describes an analysis of off-shell production of Higgs bosons in the \hstarww~ channel, using leptonic final states and performed on the $\sqrt{s}=13$\,\TeV $pp$ collision data-set collected with the ATLAS detector during Run 2 of the LHC. No significant deviation from the Standard Model expectation is observed. The data are used to constrain the signal strength for off-shell Higgs boson production, yielding an observed (expected) measurement of $\muoff$ at $0.3^{+0.9}_{-0.3}$ ($1.0^{+2.3}_{-1.0}$) with an upper bound of 3.4 (4.4) at 95\% CL. This is a significant improvement over the ATLAS LHC Run~1 measurement of $\muoff$ in the $\hwwstar$~ channel, which found an observed (expected) upper limit of 17.2 (21.3) at 95\% CL~\cite{HIGG-2014-10}.

This analysis is combined with a measurement of on-shell Higgs boson production in the $\hwwstar$~ channel, and the results of the combination are interpreted as a constraint on the Higgs boson total width. Under the assumption that the off- and on-shell coupling modifiers are the same for both ggF and EW production modes, the observed (expected) value for $\gammaH$ is $0.9^{+3.4}_{-0.9}$ ($4.1^{+8.3}_{-3.8}$)\,\MeV, where the Standard Model predicts a value of 4.1\,\MeV~\cite{deFlorian:2016spz}. An observed (expected) upper bound at $95\%$ confidence level on the Higgs boson total width is measured at 13.1 (17.3)~\MeV.

%

%

% End of text imported from the .//sections/conclusions.tex input file

%
\section*{Acknowledgements}
%

% The next lines are included from the .//acknowledgements/Acknowledgements.tex input file
%
%

%
%

We thank CERN for the very successful operation of the LHC and its injectors, as well as the support staff at
CERN and at our institutions worldwide without whom ATLAS could not be operated efficiently.

The crucial computing support from all WLCG partners is acknowledged gratefully, in particular from CERN, the ATLAS Tier-1 facilities at TRIUMF/SFU (Canada), NDGF (Denmark, Norway, Sweden), CC-IN2P3 (France), KIT/GridKA (Germany), INFN-CNAF (Italy), NL-T1 (Netherlands), PIC (Spain), RAL (UK) and BNL (USA), the Tier-2 facilities worldwide and large non-WLCG resource providers. Major contributors of computing resources are listed in Ref.~\cite{ATL-SOFT-PUB-2025-001}.

We gratefully acknowledge the support of ANPCyT, Argentina; YerPhI, Armenia; ARC, Australia; BMWFW and FWF, Austria; ANAS, Azerbaijan; CNPq and FAPESP, Brazil; NSERC, NRC and CFI, Canada; CERN; ANID, Chile; CAS, MOST and NSFC, China; Minciencias, Colombia; MEYS CR, Czech Republic; DNRF and DNSRC, Denmark; IN2P3-CNRS and CEA-DRF/IRFU, France; SRNSFG, Georgia; BMFTR, HGF and MPG, Germany; GSRI, Greece; RGC and Hong Kong SAR, China; ICHEP and Academy of Sciences and Humanities, Israel; INFN, Italy; MEXT and JSPS, Japan; CNRST, Morocco; NWO, Netherlands; RCN, Norway; MNiSW, Poland; FCT, Portugal; MNE/IFA, Romania; MSTDI, Serbia; MSSR, Slovakia; ARIS and MVZI, Slovenia; DSI/NRF, South Africa; MICIU/AEI, Spain; SRC and Wallenberg Foundation, Sweden; SERI, SNSF and Cantons of Bern and Geneva, Switzerland; NSTC, Taipei; TENMAK, T\"urkiye; STFC/UKRI, United Kingdom; DOE and NSF, United States of America.

Individual groups and members have received support from BCKDF, CANARIE, CRC and DRAC, Canada; CERN-CZ, FORTE and PRIMUS, Czech Republic; COST, ERC, ERDF, Horizon 2020, ICSC-NextGenerationEU and Marie Sk{\l}odowska-Curie Actions, European Union; Investissements d'Avenir Labex, Investissements d'Avenir Idex and ANR, France; DFG and AvH Foundation, Germany; Herakleitos, Thales and Aristeia programmes co-financed by EU-ESF and the Greek NSRF, Greece; BSF-NSF and MINERVA, Israel; NCN and NAWA, Poland; La Caixa Banking Foundation, CERCA Programme Generalitat de Catalunya and PROMETEO and GenT Programmes Generalitat Valenciana, Spain; G\"{o}ran Gustafssons Stiftelse, Sweden; The Royal Society and Leverhulme Trust, United Kingdom.

In addition, individual members wish to acknowledge support from Armenia: Yerevan Physics Institute (FAPERJ); CERN: European Organization for Nuclear Research (CERN DOCT); Chile: Agencia Nacional de Investigaci\'on y Desarrollo (FONDECYT 1230812, FONDECYT 1240864); China: Chinese Ministry of Science and Technology (MOST-2023YFA1605700, MOST-2023YFA1609300), National Natural Science Foundation of China (NSFC - 12175119, NSFC 12275265); Czech Republic: Czech Science Foundation (GACR - 24-11373S), Ministry of Education Youth and Sports (ERC-CZ-LL2327, FORTE CZ.02.01.01/00/22\_008/0004632), PRIMUS Research Programme (PRIMUS/21/SCI/017); EU: H2020 European Research Council (ERC - 101002463); European Union: European Research Council (BARD No. 101116429, ERC - 948254, ERC 101089007), European Regional Development Fund (SMASH COFUND 101081355, SLO ERDF), Horizon 2020 Framework Programme (MUCCA - CHIST-ERA-19-XAI-00), European Union, Future Artificial Intelligence Research (FAIR-NextGenerationEU PE00000013), Horizon 2020 (EuroHPC - EHPC-DEV-2024D11-051), Italian Center for High Performance Computing, Big Data and Quantum Computing (ICSC, NextGenerationEU); France: Agence Nationale de la Recherche (ANR-21-CE31-0022, ANR-22-EDIR-0002); Germany: Baden-Württemberg Stiftung (BW Stiftung-Postdoc Eliteprogramme), Deutsche Forschungsgemeinschaft (DFG - 469666862, DFG - CR 312/5-2); China: Research Grants Council (GRF); Italy: Istituto Nazionale di Fisica Nucleare (ICSC, NextGenerationEU), Ministero dell'Università e della Ricerca (NextGenEU 153D23001490006 M4C2.1.1, NextGenEU I53D23000820006 M4C2.1.1, NextGenEU I53D23001490006 M4C2.1.1); Japan: Japan Society for the Promotion of Science (JSPS KAKENHI JP22H01227, JSPS KAKENHI JP22H04944, JSPS KAKENHI JP22KK0227, JSPS KAKENHI JP23KK0245, JSPS KAKENHI JP24K23939); Norway: Research Council of Norway (RCN-314472); Poland: Ministry of Science and Higher Education (IDUB AGH, POB8, D4 no 9722), Polish National Science Centre (NCN 2021/42/E/ST2/00350, NCN OPUS 2023/51/B/ST2/02507, NCN OPUS nr 2022/47/B/ST2/03059, NCN UMO-2019/34/E/ST2/00393, UMO-2022/47/O/ST2/00148, UMO-2023/49/B/ST2/04085, UMO-2023/51/B/ST2/00920, UMO-2024/53/N/ST2/00869); Portugal: Foundation for Science and Technology (FCT); Spain: Generalitat Valenciana (Artemisa, FEDER, IDIFEDER/2018/048), Ministry of Science and Innovation (MCIN \& NextGenEU PCI2022-135018-2, MICIN \& FEDER PID2021-125273NB, RYC2019-028510-I, RYC2020-030254-I, RYC2021-031273-I, RYC2022-038164-I); Sweden: Carl Trygger Foundation (Carl Trygger Foundation CTS 22:2312), Swedish Research Council (Swedish Research Council 2023-04654, VR 2021-03651, VR 2022-03845, VR 2022-04683, VR 2023-03403, VR 2024-05451), Knut and Alice Wallenberg Foundation (KAW 2018.0458, KAW 2022.0358, KAW 2023.0366); Switzerland: Swiss National Science Foundation (SNSF - PCEFP2\_194658); United Kingdom: Leverhulme Trust (Leverhulme Trust RPG-2020-004), Royal Society (NIF-R1-231091); United States of America: U.S. Department of Energy (ECA DE-AC02-76SF00515), Neubauer Family Foundation.

%
%

% End of text imported from the .//acknowledgements/Acknowledgements.tex input file

%
%
%
%
%
%
%

%
\clearpage
\printbibliography

@article{Passarino:2012ri,
    author = "Passarino, Giampiero",
    title = "{Higgs interference effects in $gg \to ZZ$ and their uncertainty}",
    eprint = "1206.3824",
    archivePrefix = "arXiv",
    primaryClass = "hep-ph",
    doi = "10.1007/JHEP08(2012)146",
    journal = "JHEP",
    volume = "08",
    pages = "146",
    year = "2012"
}

@article{Czakon:2017wor,
    author = "Czakon, Michal and Heymes, David and Mitov, Alexander and Pagani, Davide and Tsinikos, Ioannis and Zaro, Marco",
    title = "{Top-pair production at the LHC through NNLO QCD and NLO EW}",
    eprint = "1705.04105",
    archivePrefix = "arXiv",
    primaryClass = "hep-ph",
    reportNumber = "CAVENDISH-HEP-17-07, CP3-17-12, TUM-HEP-1084-17, TTK-17-15",
    doi = "10.1007/JHEP10(2017)186",
    journal = "JHEP",
    volume = "10",
    pages = "186",
    year = "2017"
}

@article{LEHMANN2023168376,
    title = {The Fake Factor Method and its relation to the Matrix Method},
    journal = {Nucl. Instrum. Meth. A},
    volume = {1054},
    pages = {168376},
    year = {2023},
    doi = {10.1016/j.nima.2023.168376},
    author = {Konstantin Lehmann and Bernd Stelzer}
}

@article{Denner_2022,
   title="{NLO QCD and EW corrections to vector-boson scattering into $W^{+}W^{-}$ at the LHC}",
   DOI="{10.1007/jhep06(2022)098}",
   pages="{098}",
   journal="{JHEP}",
   publisher="{Springer Science and Business Media LLC}",
   author="{Denner, A. and Franken, R. and Schmidt, T. and Schwan, C.}",
   year="2022",
   month="6",
   eprint = "2202.10844",
   archivePrefix = "arXiv",
   primaryClass = "hep-ph",
   volume = "11",
}

@misc{tensorflow2015-whitepaper,
title={ {TensorFlow}: Large-Scale Machine Learning on Heterogeneous Systems},
url={https://www.tensorflow.org/},
note={Software available from tensorflow.org},
author={
    Mart\'{i}n~Abadi and
    Ashish~Agarwal and
    Paul~Barham and
    Eugene~Brevdo and
    Zhifeng~Chen and
    Craig~Citro and
    Greg~S.~Corrado and
    Andy~Davis and
    Jeffrey~Dean and
    Matthieu~Devin and
    Sanjay~Ghemawat and
    Ian~Goodfellow and
    Andrew~Harp and
    Geoffrey~Irving and
    Michael~Isard and
    Yangqing Jia and
    Rafal~Jozefowicz and
    Lukasz~Kaiser and
    Manjunath~Kudlur and
    Josh~Levenberg and
    Dandelion~Man\'{e} and
    Rajat~Monga and
    Sherry~Moore and
    Derek~Murray and
    Chris~Olah and
    Mike~Schuster and
    Jonathon~Shlens and
    Benoit~Steiner and
    Ilya~Sutskever and
    Kunal~Talwar and
    Paul~Tucker and
    Vincent~Vanhoucke and
    Vijay~Vasudevan and
    Fernanda~Vi\'{e}gas and
    Oriol~Vinyals and
    Pete~Warden and
    Martin~Wattenberg and
    Martin~Wicke and
    Yuan~Yu and
    Xiaoqiang~Zheng},
  year={2015},
}

@article{Anastasiou:2005qj,
    author = "Anastasiou, Charalampos and Melnikov, Kirill and Petriello, Frank",
    title = "{Fully differential Higgs boson production and the di-photon signal through next-to-next-to-leading order}",
    eprint = "hep-ph/0501130",
    archivePrefix = "arXiv",
    reportNumber = "UH-511-1066-05",
    doi = "10.1016/j.nuclphysb.2005.06.036",
    journal = "Nucl. Phys. B",
    volume = "724",
    pages = "197--246",
    year = "2005"
}

@article{Grazzini:2017mhc,
    author = "Grazzini, Massimiliano and Kallweit, Stefan and Wiesemann, Marius",
    title = "{Fully differential NNLO computations with MATRIX}",
    eprint = "1711.06631",
    archivePrefix = "arXiv",
    primaryClass = "hep-ph",
    reportNumber = "ZU-TH-30-17, CERN-TH-2017-232, ZU-TH 30/17",
    doi = "10.1140/epjc/s10052-018-5771-7",
    journal = "Eur. Phys. J. C",
    volume = "78",
    number = "7",
    pages = "537",
    year = "2018"
}

@inproceedings{kohavi1995study,
  title="{A study of cross-validation and bootstrap for accuracy estimation and model selection}",
  author={Kohavi, Ron and others},
  booktitle={Ijcai},
  volume={14},
  number={2},
  pages={1137--1145},
  year={1995},
  url = {http://dblp.uni-trier.de/db/conf/ijcai/ijcai95.html#Kohavi95},
  organization={Montreal, Canada}
}

@article{farina2012dys,
  title={(Dys) Zphilia or a custodial breaking Higgs at the LHC},
  author={Farina, Marco and Grojean, Christophe and Salvioni, Ennio},
  journal = "JHEP",
  volume={2012},
  number={12},
  pages={1--18},
  year={2012},
  eprint = "1205.0011",
  archivePrefix = "arXiv",
  primaryClass = "hep-ph",
  doi="10.1007/JHEP07(2012)012"
}

@article{butterworth2024custodial,
  title={Custodial symmetry breaking and Higgs boson signatures at the LHC},
  author={Butterworth, Jon and Debnath, Hridoy and Fileviez P{\'e}rez, Pavel and Mitchell, Francis},
  journal        = "Phys. Rev. D",
  volume={109},
  number={9},
  pages={095014},
  year={2024},
  eprint = "2309.10027",
  archivePrefix = "arXiv",
  primaryClass = "hep-ph",
  doi="10.1103/PhysRevD.109.095014"
}

@article{solberg2013terms,
  title={On the terms violating the custodial symmetry in multi-Higgs-doublet models},
  author={Solberg, M Aa},
  journal={J. Phys. G},
  volume={40},
  number={6},
  pages={065001},
  year={2013},
  eprint = "1207.5194",
  archivePrefix = "arXiv",
  primaryClass = "hep-ph",
  doi="10.1088/0954-3899/40/6/065001"
}

@Article{Evans:2008zzb,
      author         = "Evans, Lyndon and Bryant, Philip",
      title          = "{LHC Machine}",
      journal        = "JINST",
      volume         = "3",
      pages          = "S08001",
      doi            = "10.1088/1748-0221/3/08/S08001",
      year           = "2008",
      SLACcitation   = "%%CITATION = JINST,3,S08001;%%",
}

@Article{Cacciari:2008gp,
     author    = "Cacciari, Matteo and Salam, Gavin P. and Soyez, Gregory",
     title     = "{The anti-\(k_{t}\) jet clustering algorithm}",
     journal   = "JHEP",
     volume    = "04",
     year      = "2008",
     pages     = "063",
     eprint    = "0802.1189",
     archivePrefix = "arXiv",
     primaryClass  =  "hep-ph",
     doi       = "10.1088/1126-6708/2008/04/063",
     SLACcitation  = "%%CITATION = 0802.1189;%%"
}

@Article{Fastjet,
      author         = "Cacciari, Matteo and Salam, Gavin P. and Soyez, Gregory",
      title          = "{FastJet user manual}",
      journal        = "Eur. Phys. J. C",
      volume         = "72",
      year           = "2012",
      pages          = "1896",
      doi            = "10.1140/epjc/s10052-012-1896-2",
      eprint         = "1111.6097",
      archivePrefix  = "arXiv",
      primaryClass   = "hep-ph",
      reportNumber   = "CERN-PH-TH-2011-297",
      SLACcitation   = "%%CITATION = ARXIV:1111.6097;%%"
}

@Article{Butterworth:2015oua,
      author         = "Butterworth, Jon and others",
      title          = "{PDF4LHC recommendations for LHC Run II}",
      journal        = "J. Phys. G",
      volume         = "43",
      year           = "2016",
      pages          = "023001",
      doi            = "10.1088/0954-3899/43/2/023001",
      eprint         = "1510.03865",
      archivePrefix  = "arXiv",
      primaryClass   = "hep-ph",
      reportNumber   = "OUTP-15-17P, SMU-HEP-15-12, TIF-UNIMI-2015-14,
                        LCTS-2015-27, CERN-PH-TH-2015-249",
      SLACcitation   = "%%CITATION = ARXIV:1510.03865;%%"
}

@article{Harland-Lang:2014zoa,
      author         = "Harland-Lang, L. A. and Martin, A. D. and Motylinski, P.
                        and Thorne, R. S.",
      title          = "{Parton distributions in the LHC era: MMHT 2014 PDFs}",
      journal        = "Eur. Phys. J. C",
      number         = "5",
      volume         = "75",
      pages          = "204",
      doi            = "10.1140/epjc/s10052-015-3397-6",
      year           = "2015",
      eprint         = "1412.3989",
      archivePrefix  = "arXiv",
      primaryClass   = "hep-ph",
      reportNumber   = "LCTS-2014-47, IPPP-14-97, DCPT-14-194",
      SLACcitation   = "%%CITATION = ARXIV:1412.3989;%%",
}

@article{Ball:2012cx,
      author         = "{NNPDF Collaboration} and Ball, Richard D. and others",
      title          = "{Parton distributions with LHC data}",
      journal        = "Nucl. Phys. B",
      volume         = "867",
      year           = "2013",
      pages          = "244",
      doi            = "10.1016/j.nuclphysb.2012.10.003",
      eprint         = "1207.1303",
      archivePrefix  = "arXiv",
      primaryClass   = "hep-ph",
      reportNumber   = "EDINBURGH-2012-08, IFUM-FT-997, FR-PHENO-2012-014,
                        RWTH-TTK-12-25, CERN-PH-TH-2012-037, SFB-CPP-12-47\,
                        --CERN-PH-TH-2012-037",
      SLACcitation   = "%%CITATION = ARXIV:1207.1303;%%"
}

@article{Ball:2014uwa,
      author         = "{NNPDF Collaboration} and Ball, Richard D. and others",
      title          = "{Parton distributions for the LHC run II}",
      journal        = "JHEP",
      volume         = "04",
      year           = "2015",
      pages          = "040",
      doi            = "10.1007/JHEP04(2015)040",
      eprint         = "1410.8849",
      archivePrefix  = "arXiv",
      primaryClass   = "hep-ph",
      reportNumber   = "EDINBURGH-2014-15, IFUM-1034-FT, CERN-PH-TH-2013-253,
                        OUTP-14-11P, CAVENDISH-HEP-14-11",
      SLACcitation   = "%%CITATION = ARXIV:1410.8849;%%"
}

@article{Czakon:2012pz,
      author         = "Czakon, Michal and Mitov, Alexander",
      title          = "{NNLO corrections to top pair production at hadron
                        colliders: the quark-gluon reaction}",
      journal        = "JHEP",
      volume         = "01",
      pages          = "080",
      doi            = "10.1007/JHEP01(2013)080",
      year           = "2013",
      eprint         = "1210.6832",
      archivePrefix  = "arXiv",
      primaryClass   = "hep-ph",
      SLACcitation   = "%%CITATION = ARXIV:1210.6832;%%",
}

@Article{Sjostrand:2007gs,
	author    = "Sj{\"o}strand, T. and Mrenna, S. and Skands, P.",
	title     = "{A brief introduction to PYTHIA 8.1}",
	journal   = "Comput. Phys. Commun.",
	volume    = "178",
	year      = "2008",
	pages     = "852-867",
	eprint    = "0710.3820",
	archivePrefix = "arXiv",
	primaryClass  =  "hep-ph",
	doi       = "10.1016/j.cpc.2008.01.036",
	SLACcitation  = "%%CITATION = 0710.3820;%%"
}

@article{Sjostrand:2014zea,
      author         = "Sj{\"o}strand, Torbj{\"o}rn and Ask, Stefan and Christiansen,
                        Jesper R. and Corke, Richard and Desai, Nishita and Ilten,
                        Philip and Mrenna, Stephen and Prestel, Stefan and
                        Rasmussen, Christine O. and Skands, Peter Z.",
      title          = "{An introduction to PYTHIA 8.2}",
      journal        = "Comput. Phys. Commun.",
      volume         = "191",
      year           = "2015",
      pages          = "159",
      doi            = "10.1016/j.cpc.2015.01.024",
      eprint         = "1410.3012",
      archivePrefix  = "arXiv",
      primaryClass   = "hep-ph",
      reportNumber   = "LU-TP-14-36, MCNET-14-22, CERN-PH-TH-2014-190,
                        FERMILAB-PUB-14-316-CD, DESY-14-178, SLAC-PUB-16122,
                        --FERMILAB-PUB-14-316-CD",
      SLACcitation   = "%%CITATION = ARXIV:1410.3012;%%"
}

@Article{Frixione:2008yi,
     author    = "Frixione, Stefano and Laenen, Eric and Motylinski, Patrick and White, Chris and Webber, Bryan R.",
     title     = "{Single-top hadroproduction in association with a \(W\) boson}",
     journal   = "JHEP",
     volume    = "07",
     year      = "2008",
     pages     = "029",
     eprint    = "0805.3067",
     archivePrefix = "arXiv",
     primaryClass  =  "hep-ph",
     doi       = "10.1088/1126-6708/2008/07/029"
}

@Article{Alwall:2014hca,
      author         = "Alwall, J. and Frederix, R. and Frixione, S. and Hirschi,
                        V. and Maltoni, F. and Mattelaer, O. and Shao, H. -S. and
                        Stelzer, T. and Torrielli, P. and Zaro, M.",
      title          = "{The automated computation of tree-level and
                        next-to-leading order differential cross sections, and
                        their matching to parton shower simulations}",
      journal        = "JHEP",
      volume         = "07",
      year           = "2014",
      pages          = "079",
      doi            = "10.1007/JHEP07(2014)079",
      eprint         = "1405.0301",
      archivePrefix  = "arXiv",
      primaryClass   = "hep-ph",
      reportNumber   = "CERN-PH-TH-2014-064, CP3-14-18, LPN14-066, MCNET-14-09,
                        ZU-TH-14-14",
      SLACcitation   = "%%CITATION = ARXIV:1405.0301;%%"
}

@Article{Bahr:2008pv,
      author         = "B{\"a}hr, M. and others",
      title          = "{Herwig++ physics and manual}",
      journal        = "Eur. Phys. J. C",
      volume         = "58",
      year           = "2008",
      pages          = "639",
      doi            = "10.1140/epjc/s10052-008-0798-9",
      eprint         = "0803.0883",
      archivePrefix  = "arXiv",
      primaryClass   = "hep-ph",
      reportNumber   = "CERN-PH-TH-2008-038, CAVENDISH-HEP-08-03, KA-TP-05-2008,
                        DCPT-08-22, IPPP-08-11, CP3-08-05",
      SLACcitation   = "%%CITATION = ARXIV:0803.0883;%%"
}

@Article{Bellm:2015jjp,
      author         = "Bellm, Johannes and others",
      title          = "{Herwig 7.0/Herwig++ 3.0 release note}",
      journal        = "Eur. Phys. J. C",
      volume         = "76",
      year           = "2016",
      number         = "4",
      pages          = "196",
      doi            = "10.1140/epjc/s10052-016-4018-8",
      eprint         = "1512.01178",
      archivePrefix  = "arXiv",
      primaryClass   = "hep-ph",
      reportNumber   = "CERN-PH-TH-2015-289, MAN-HEP-2015-15, IFJPAN-IV-2015-13,
                        HERWIG-2015-01, KA-TP-18-2015, DCPT-15-142, MCNET-15-28,
                        IPPP-15-71, --HERWIG-2015-01",
      SLACcitation   = "%%CITATION = ARXIV:1512.01178;%%"
}

@Article{Nason:2004rx,
      author         = "Nason, Paolo",
      title          = "{A new method for combining NLO QCD with shower Monte Carlo algorithms}",
      journal        = "JHEP",
      volume         = "11",
      pages          = "040",
      doi            = "10.1088/1126-6708/2004/11/040",
      year           = "2004",
      eprint         = "hep-ph/0409146",
      archivePrefix  = "arXiv",
}

@Article{Frixione:2007vw,
      author         = "Frixione, Stefano and Nason, Paolo and Oleari, Carlo",
      title          = "{Matching NLO QCD computations with parton shower
                        simulations: the POWHEG method}",
      journal        = "JHEP",
      volume         = "11",
      pages          = "070",
      doi            = "10.1088/1126-6708/2007/11/070",
      year           = "2007",
      eprint         = "0709.2092",
      archivePrefix  = "arXiv",
      primaryClass   = "hep-ph",
}

@Article{Alioli:2010xd,
      author         = "Alioli, Simone and Nason, Paolo and Oleari, Carlo and Re,
                        Emanuele",
      title          = "{A general framework for implementing NLO calculations in
                        shower Monte Carlo programs: the POWHEG BOX}",
      journal        = "JHEP",
      volume         = "06",
      pages          = "043",
      doi            = "10.1007/JHEP06(2010)043",
      year           = "2010",
      eprint         = "1002.2581",
      archivePrefix  = "arXiv",
      primaryClass   = "hep-ph",
}

@article{Hartanto:2015uka,
      author         = "Hartanto, Heribertus B. and J{\"a}ger, Barbara and Reina,
                        Laura and Wackeroth, Doreen",
      title          = "{Higgs boson production in association with top quarks in
                        the POWHEG BOX}",
      journal        = "Phys. Rev. D",
      volume         = "91",
      year           = "2015",
      number         = "9",
      pages          = "094003",
      doi            = "10.1103/PhysRevD.91.094003",
      eprint         = "1501.04498",
      archivePrefix  = "arXiv",
      primaryClass   = "hep-ph",
      SLACcitation   = "%%CITATION = ARXIV:1501.04498;%%"
}

@article{Re:2010bp,
      author         = "Re, Emanuele",
      title          = "{Single-top \(Wt\)-channel production matched with parton
                        showers using the POWHEG method}",
      journal        = "Eur. Phys. J. C",
      volume         = "71",
      year           = "2011",
      pages          = "1547",
      doi            = "10.1140/epjc/s10052-011-1547-z",
      eprint         = "1009.2450",
      archivePrefix  = "arXiv",
      primaryClass   = "hep-ph",
      reportNumber   = "IPPP-10-74, DCPT-10-148",
      SLACcitation   = "%%CITATION = ARXIV:1009.2450;%%"
}

@article{Kidonakis:2010ux,
      author         = "Kidonakis, Nikolaos",
      title          = "{Two-loop soft anomalous dimensions for single top quark
                        associated production with a \(W^{-}\) or \(H^{-}\)}",
      journal        = "Phys. Rev. D",
      volume         = "82",
      year           = "2010",
      pages          = "054018",
      doi            = "10.1103/PhysRevD.82.054018",
      eprint         = "1005.4451",
      archivePrefix  = "arXiv",
      primaryClass   = "hep-ph",
      SLACcitation   = "%%CITATION = ARXIV:1005.4451;%%"
}

@inproceedings{Kidonakis:2013zqa,
      author         = "Kidonakis, Nikolaos",
      title          = "{Top Quark Production}",
      booktitle      = "{Proceedings, Helmholtz International Summer School on Physics of Heavy Quarks and Hadrons (HQ2013)}",
      eventdate      = "2013-07-15/2013-07-28",
      venue          = "JINR, Dubna, Russia",
      pages          = "139-168",
      doi            = "10.3204/DESY-PROC-2013-03/Kidonakis",
      eprint         = "1311.0283",
      archivePrefix  = "arXiv",
      primaryClass   = "hep-ph",
      reportNumber   = "KSU-HEP-110113",
      SLACcitation   = "%%CITATION = ARXIV:1311.0283;%%"
}

@article{Beneke:2011mq,
      author         = "Beneke, M. and Falgari, P. and Klein, S. and Schwinn, C.",
      title          = "{Hadronic top-quark pair production with NNLL threshold
                        resummation}",
      journal        = "Nucl. Phys. B",
      volume         = "855",
      year           = "2012",
      pages          = "695-741",
      doi            = "10.1016/j.nuclphysb.2011.10.021",
      eprint         = "1109.1536",
      archivePrefix  = "arXiv",
      primaryClass   = "hep-ph",
      reportNumber   = "TTK-11-38, ITP-UU-11-26, SPIN-11-19, FR-PHENO-2011-015,
                        SFB-CPP-11-49",
      SLACcitation   = "%%CITATION = ARXIV:1109.1536;%%"
}

@article{Anastasiou:2003ds,
      author         = "Anastasiou, Charalampos and Dixon, Lance and Melnikov,
                        Kirill and Petriello, Frank",
      title          = "{High-precision QCD at hadron colliders: Electroweak
                        gauge boson rapidity distributions at next-to-next-to leading order}",
      journal        = "Phys. Rev. D",
      volume         = "69",
      year           = "2004",
      pages          = "094008",
      doi            = "10.1103/PhysRevD.69.094008",
      eprint         = "hep-ph/0312266",
      archivePrefix  = "arXiv",
      reportNumber   = "SLAC-PUB-10288, UH-511-1042-03",
      SLACcitation   = "%%CITATION = HEP-PH/0312266;%%"
}

@article{Cacciari:2011hy,
      author         = "Cacciari, Matteo and Czakon, Michal and Mangano,
                        Michelangelo and Mitov, Alexander and Nason, Paolo",
      title          = "{Top-pair production at hadron colliders with
                        next-to-next-to-leading logarithmic soft-gluon
                        resummation}",
      journal        = "Phys. Lett. B",
      volume         = "710",
      pages          = "612-622",
      doi            = "10.1016/j.physletb.2012.03.013",
      year           = "2012",
      eprint         = "1111.5869",
      primaryClass   = "hep-ph",
      archivePrefix  = "arXiv",
      reportNumber   = "CERN-PH-TH-2011-277, TTK-11-54",
      SLACcitation   = "%%CITATION = ARXIV:1111.5869;%%",
}

@article{Czakon:2012zr,
      author         = "Czakon, Michal and Mitov, Alexander",
      title          = "{NNLO corrections to top-pair production at hadron
                        colliders: the all-fermionic scattering channels}",
      journal        = "JHEP",
      volume         = "12",
      pages          = "054",
      doi            = "10.1007/JHEP12(2012)054",
      year           = "2012",
      eprint         = "1207.0236",
      archivePrefix  = "arXiv",
      primaryClass   = "hep-ph",
      SLACcitation   = "%%CITATION = ARXIV:1207.0236;%%",
}

@article{Czakon:2013goa,
      author         = "Czakon, Michal and Fiedler, Paul and Mitov, Alexander",
      title          = "{Total Top-Quark Pair-Production Cross Section at
                        Hadron Colliders Through \(\mathcal{O}(\alpha_S^4)\)}",
      journal        = "Phys. Rev. Lett.",
      volume         = "110",
      pages          = "252004",
      doi            = "10.1103/PhysRevLett.110.252004",
      year           = "2013",
      eprint         = "1303.6254",
      archivePrefix  = "arXiv",
      primaryClass   = "hep-ph",
      reportNumber   = "CERN-PH-TH-2013-056, TTK-13-08",
      SLACcitation   = "%%CITATION = ARXIV:1303.6254;%%",
}

@article{Baernreuther:2012ws,
      author         = "B{\"a}rnreuther, Peter and Czakon, Michal and Mitov,
                        Alexander",
      title          = "{Percent-Level-Precision Physics at the
                  Tevatron: Next-to-Next-to-Leading Order QCD
                  Corrections to \(q \bar{q} \to t \bar{t} + X\)}",
      journal        = "Phys. Rev. Lett.",
      volume         = "109",
      pages          = "132001",
      doi            = "10.1103/PhysRevLett.109.132001",
      year           = "2012",
      eprint         = "1204.5201",
      archivePrefix  = "arXiv",
      primaryClass   = "hep-ph",
      SLACcitation   = "%%CITATION = ARXIV:1204.5201;%%",
}

@article{Frixione:2007nw,
      author         = "Frixione, Stefano and Ridolfi, Giovanni and Nason, Paolo",
      title          = "{A positive-weight next-to-leading-order Monte Carlo for
                        heavy flavour hadroproduction}",
      journal        = "JHEP",
      volume         = "09",
      pages          = "126",
      doi            = "10.1088/1126-6708/2007/09/126",
      year           = "2007",
      eprint         = "0707.3088",
      archivePrefix  = "arXiv",
      primaryClass   = "hep-ph",
      SLACcitation   = "%%CITATION = ARXIV:0707.3088;%%",
}

@article{deFlorian:2016spz,
      author         = "de Florian, D. and others",
      title          = "{Handbook of LHC Higgs Cross Sections: 4. Deciphering the Nature of the Higgs Sector}",
      collaboration  = "LHC Higgs Cross Section Working Group",
      doi            = "10.23731/CYRM-2017-002",
      year           = "2017",
      eprint         = "1610.07922",
      archivePrefix  = "arXiv",
      primaryClass   = "hep-ph",
      reportNumber   = "FERMILAB-FN-1025-T, CERN-2017-002-M",
      SLACcitation   = "%%CITATION = ARXIV:1610.07922;%%"
}

@Article{Hoeche:2009rj,
      author         = "H{\"o}che, Stefan and Krauss, Frank and Schumann, Steffen and Siegert, Frank",
      title          = "{QCD matrix elements and truncated showers}",
      journal        = "JHEP",
      volume         = "05",
      pages          = "053",
      doi            = "10.1088/1126-6708/2009/05/053",
      year           = "2009",
      eprint         = "0903.1219",
      archivePrefix  = "arXiv",
      primaryClass   = "hep-ph",
}

@Article{Gleisberg:2008fv,
      author         = "Gleisberg, Tanju and H{\"o}che, Stefan",
      title          = "{Comix, a new matrix element generator}",
      journal        = "JHEP",
      volume         = "12",
      pages          = "039",
      doi            = "10.1088/1126-6708/2008/12/039",
      year           = "2008",
      eprint         = "0808.3674",
      archivePrefix  = "arXiv",
      primaryClass   = "hep-ph",
}

@Article{Schumann:2007mg,
      author         = "Schumann, Steffen and Krauss, Frank",
      title          = "{A parton shower algorithm based on Catani--Seymour dipole factorisation}",
      journal        = "JHEP",
      volume         = "03",
      pages          = "038",
      doi            = "10.1088/1126-6708/2008/03/038",
      year           = "2008",
      eprint         = "0709.1027",
      archivePrefix  = "arXiv",
      primaryClass   = "hep-ph",
}

@article{Hoeche:2012yf,
      author         = "H{\"o}che, Stefan and Krauss, Frank and Sch{\"o}nherr, Marek and
                        Siegert, Frank",
      title          = "{QCD matrix elements + parton showers. The NLO case}",
      journal        = "JHEP",
      volume         = "04",
      year           = "2013",
      pages          = "027",
      doi            = "10.1007/JHEP04(2013)027",
      eprint         = "1207.5030",
      archivePrefix  = "arXiv",
      primaryClass   = "hep-ph",
      reportNumber   = "SLAC-PUB-15191, IPPP-12-52, DCPT-12-104, LPN12-081,
                        FR-PHENO-2012-017, MCNET-12-09, --FR-PHENO-2012-017",
      SLACcitation   = "%%CITATION = ARXIV:1207.5030;%%"
}

@article{Bothmann:2019yzt,
      author = "Bothmann, Enrico and others",
      title = "{Event generation with Sherpa 2.2}",
      journal = "SciPost Phys.",
      volume = "7",
      year = "2019",
      number = "3",
      pages = "034",
      doi = "10.21468/SciPostPhys.7.3.034",
      reportNumber = "FERMILAB-PUB-19-218-T, SLAC-PUB-17433, IPPP/19/42, MCNET-19-11",
      eprint = "1905.09127",
      archivePrefix = "arXiv",
      primaryClass = "hep-ph",
}

@article{Hoeche:2011fd,
      author         = "H{\"o}che, Stefan and Krauss, Frank and Sch{\"o}nherr, Marek and
                        Siegert, Frank",
      title          = "{A critical appraisal of NLO+PS matching methods}",
      journal        = "JHEP",
      volume         = "09",
      year           = "2012",
      pages          = "049",
      doi            = "10.1007/JHEP09(2012)049",
      eprint         = "1111.1220",
      archivePrefix  = "arXiv",
      primaryClass   = "hep-ph",
      reportNumber   = "SLAC-PUB-14661, IPPP-11-67, DCPT-11-134, LPN11-58,
                        FR-PHENO-2011-019, MCNET-11-24",
      SLACcitation   = "%%CITATION = ARXIV:1111.1220;%%"
}

@article{Catani:2001cc,
      author         = "Catani, S. and Krauss, F. and Webber, B. R. and Kuhn, R.",
      title          = "{QCD Matrix Elements + Parton Showers}",
      journal        = "JHEP",
      volume         = "11",
      year           = "2001",
      pages          = "063",
      doi            = "10.1088/1126-6708/2001/11/063",
      eprint         = "hep-ph/0109231",
      archivePrefix  = "arXiv",
      reportNumber   = "CERN-TH-2000-367, CAVENDISH-HEP-00-03",
      SLACcitation   = "%%CITATION = HEP-PH/0109231;%%"
}

@article{Cascioli:2011va,
      author         = "Cascioli, Fabio and Maierh{\"o}fer, Philipp and Pozzorini, Stefano",
      title          = "{Scattering Amplitudes with Open Loops}",
      journal        = "Phys. Rev. Lett.",
      volume         = "108",
      year           = "2012",
      pages          = "111601",
      doi            = "10.1103/PhysRevLett.108.111601",
      eprint         = "1111.5206",
      archivePrefix  = "arXiv",
      primaryClass   = "hep-ph",
      reportNumber   = "ZU-TH-23-11, LPN11-66",
      SLACcitation   = "%%CITATION = ARXIV:1111.5206;%%"
}

@article{Buccioni:2017yxi,
    author = "Buccioni, Federico and Pozzorini, Stefano and Zoller, Max",
    title = "{On-the-fly reduction of open loops}",
    eprint = "1710.11452",
    archivePrefix = "arXiv",
    primaryClass = "hep-ph",
    reportNumber = "ZU-TH-29-17",
    doi = "10.1140/epjc/s10052-018-5562-1",
    journal = "Eur. Phys. J. C",
    volume = "78",
    number = "1",
    pages = "70",
    year = "2018"
}

@article{Buccioni:2019sur,
    author = "Buccioni, Federico and Lang, Jean-Nicolas and Lindert, Jonas M. and Maierh{\"o}fer, Philipp and Pozzorini, Stefano and Zhang, Hantian and Zoller, Max F.",
    title = "{OpenLoops 2}",
    eprint = "1907.13071",
    archivePrefix = "arXiv",
    primaryClass = "hep-ph",
    reportNumber = "IPPP/19/62, FR-PHENO-2019-12, PSI-PR-19-15, ZU-TH 37/19",
    doi = "10.1140/epjc/s10052-019-7306-2",
    journal = "Eur. Phys. J. C",
    volume = "79",
    number = "10",
    pages = "866",
    year = "2019"
}

@article{Denner:2016kdg,
      author         = "Denner, Ansgar and Dittmaier, Stefan and Hofer, Lars",
      title          = "{\textsc{Collier}: A fortran-based complex one-loop library in
                        extended regularizations}",
      journal        = "Comput. Phys. Commun.",
      volume         = "212",
      year           = "2017",
      pages          = "220-238",
      doi            = "10.1016/j.cpc.2016.10.013",
      eprint         = "1604.06792",
      archivePrefix  = "arXiv",
      primaryClass   = "hep-ph",
      reportNumber   = "FR-PHENO-2016-003, ICCUB-16-016",
      SLACcitation   = "%%CITATION = ARXIV:1604.06792;%%"
}

@article{Kallweit:2014xda,
    author = "Kallweit, Stefan and Lindert, Jonas M. and Maierh{\"o}fer, Philipp and Pozzorini, Stefano and Sch{\"o}nherr, Marek",
    title = "{NLO electroweak automation and precise predictions for W+multijet production at the LHC}",
    eprint = "1412.5157",
    archivePrefix = "arXiv",
    primaryClass = "hep-ph",
    reportNumber = "LPN14-127, IPPP-14-107, DCPT-14-214, MCNET-14-26, ZU-TH-42-14, MITP-14-102",
    doi = "10.1007/JHEP04(2015)012",
    journal = "JHEP",
    volume = "04",
    pages = "012",
    year = "2015"
}

@article{Czakon:2011xx,
  author         = "Czakon, Michal and Mitov, Alexander",
  title          = "{Top++: A program for the calculation of the top-pair cross-section at hadron colliders}",
  journal        = "Comput. Phys. Commun.",
  volume         = "185",
  year           = "2014",
  pages          = "2930",
  doi            = "10.1016/j.cpc.2014.06.021",
  eprint         = "1112.5675",
  archivePrefix  = "arXiv",
  primaryClass   = "hep-ph",
  reportNumber   = "CERN-PH-TH-2011-303, TTK-11-58",
}

@article{Catani:2007vq,
      author         = "Catani, Stefano and Grazzini, Massimiliano",
      title          = "{Next-to-Next-to-Leading-Order Subtraction Formalism in Hadron Collisions and
                        its Application to Higgs-Boson Production at the Large Hadron Collider}",
      journal        = "Phys. Rev. Lett.",
      volume         = "98",
      year           = "2007",
      pages          = "222002",
      doi            = "10.1103/PhysRevLett.98.222002",
      eprint         = "hep-ph/0703012",
      archivePrefix  = "arXiv",
      SLACcitation   = "%%CITATION = HEP-PH/0703012;%%"
}

@article{Nason:2009ai,
      author         = "Nason, Paolo and Oleari, Carlo",
      title          = "{NLO Higgs boson production via vector-boson fusion
                        matched with shower in POWHEG}",
      journal        = "JHEP",
      volume         = "02",
      year           = "2010",
      pages          = "037",
      doi            = "10.1007/JHEP02(2010)037",
      eprint         = "0911.5299",
      archivePrefix  = "arXiv",
      primaryClass   = "hep-ph",
      SLACcitation   = "%%CITATION = ARXIV:0911.5299;%%"
}

@article{Anastasiou:2015ema,
      author         = "Anastasiou, Charalampos and Duhr, Claude and Dulat, Falko
                        and Herzog, Franz and Mistlberger, Bernhard",
      title          = "{Higgs Boson Gluon-Fusion Production in QCD at Three
                        Loops}",
      journal        = "Phys. Rev. Lett.",
      volume         = "114",
      year           = "2015",
      pages          = "212001",
      doi            = "10.1103/PhysRevLett.114.212001",
      eprint         = "1503.06056",
      archivePrefix  = "arXiv",
      primaryClass   = "hep-ph",
      reportNumber   = "CERN-PH-TH-2015-055, CP3-15-07",
      SLACcitation   = "%%CITATION = ARXIV:1503.06056;%%"
}

@article{Luisoni:2013cuh,
    author = "Luisoni, Gionata and Nason, Paolo and Oleari, Carlo and Tramontano, Francesco",
    title = "{\(HW^{\pm}/HZ\) + 0 and 1 jet at NLO with the POWHEG BOX interfaced to GoSam and their merging within MiNLO}",
    eprint = "1306.2542",
    archivePrefix = "arXiv",
    primaryClass = "hep-ph",
    doi = "10.1007/JHEP10(2013)083",
    journal = "JHEP",
    volume = "10",
    pages = "083",
    year = "2013"
}

@article{Agostinelli:2002hh,
      author         = "Agostinelli, S. and others",
      title          = "{\textsc{Geant4} -- a simulation toolkit}",
      journal        = "Nucl. Instrum. Meth. A",
      volume         = "506",
      year           = "2003",
      pages          = "250",
      doi            = "10.1016/S0168-9002(03)01368-8",
      reportNumber   = "SLAC-PUB-9350, FERMILAB-PUB-03-339",
      SLACcitation   = "%%CITATION = NUIMA,A506,250;%%"
}

@Article{Cowan:2010js,
   Author = {Cowan, Glen and Cranmer, Kyle and Gross, Eilam and Vitells, Ofer},
   Title = {{Asymptotic formulae for likelihood-based tests of new physics}},
   Journal = {Eur. Phys. J. C},
   Volume = {71},
   Year = {2011},
   Pages = {1554},
   doi = "10.1140/epjc/s10052-011-1554-0",
   Eprint = "1007.1727",
   Archiveprefix = {arXiv},
   Primaryclass = {physics.data-an},
   related = "Cowan:2010js-err",
   relatedstring  = "Erratum:",
}

@Article{PERF-2007-01,
    author         = "{ATLAS Collaboration}",
    title          = "{The ATLAS Experiment at the CERN Large Hadron Collider}",
    journal        = "JINST",
    volume         = "3",
    year           = "2008",
    pages          = "S08003",
    doi            = "10.1088/1748-0221/3/08/S08003",
    primaryClass   = "hep-ex",
}

@Article{SOFT-2010-01,
    author         = "{ATLAS Collaboration}",
    title          = "{The ATLAS Simulation Infrastructure}",
    journal        = "Eur. Phys. J. C",
    volume         = "70",
    year           = "2010",
    pages          = "823",
    doi            = "10.1140/epjc/s10052-010-1429-9",
    eprint         = "1005.4568",
    archivePrefix  = "arXiv",
    primaryClass   = "physics.ins-det",
}

@Article{HIGG-2012-27,
    author         = "{ATLAS Collaboration}",
    title          = "{Observation of a new particle in the search for the Standard Model Higgs boson with the ATLAS detector at the LHC}",
    journal        = "Phys. Lett. B",
    volume         = "716",
    year           = "2012",
    pages          = "1",
    doi            = "10.1016/j.physletb.2012.08.020",
    reportNumber   = "CERN-PH-EP-2012-218",
    eprint         = "1207.7214",
    archivePrefix  = "arXiv",
    primaryClass   = "hep-ex",
}

@Article{STDM-2012-23,
    author         = "{ATLAS Collaboration}",
    title          = "{Measurement of the \(Z/\gamma^*\) boson transverse momentum distribution in \(pp\) collisions at \(\sqrt{s} = 7\,\text{TeV}\) with the ATLAS detector}",
    journal        = "JHEP",
    volume         = "09",
    year           = "2014",
    pages          = "145",
    doi            = "10.1007/JHEP09(2014)145",
    reportNumber   = "CERN-PH-EP-2014-075",
    eprint         = "1406.3660",
    archivePrefix  = "arXiv",
    primaryClass   = "hep-ex",
}

@Article{HIGG-2014-10,
    author         = "{ATLAS Collaboration}",
    title          = "{Constraints on the off-shell Higgs boson signal strength in the high-mass \(ZZ\) and \(WW\) final states with the ATLAS detector}",
    journal        = "Eur. Phys. J. C",
    volume         = "75",
    year           = "2015",
    pages          = "335",
    doi            = "10.1140/epjc/s10052-015-3542-2",
    reportNumber   = "CERN-PH-EP-2015-026",
    eprint         = "1503.01060",
    archivePrefix  = "arXiv",
    primaryClass   = "hep-ex",
}

@Article{HIGG-2014-14,
    author         = "{ATLAS and CMS Collaborations}",
    title          = "{Combined Measurement of the Higgs Boson Mass in \(pp\) Collisions at \(\sqrt{s} = 7\) and \(8\,\text{TeV}\) with the ATLAS and CMS Experiments}",
    journal        = "Phys. Rev. Lett.",
    volume         = "114",
    year           = "2015",
    pages          = "191803",
    doi            = "10.1103/PhysRevLett.114.191803",
    reportNumber   = "CERN-PH-EP-2015-075",
    eprint         = "1503.07589",
    archivePrefix  = "arXiv",
    primaryClass   = "hep-ex",
}

@Article{PERF-2015-09,
    author         = "{ATLAS Collaboration}",
    title          = "{Jet reconstruction and performance using particle flow with the ATLAS Detector}",
    journal        = "Eur. Phys. J. C",
    volume         = "77",
    year           = "2017",
    pages          = "466",
    doi            = "10.1140/epjc/s10052-017-5031-2",
    reportNumber   = "CERN-EP-2017-024",
    eprint         = "1703.10485",
    archivePrefix  = "arXiv",
    primaryClass   = "hep-ex",
}

@Article{PERF-2016-04,
    author         = "{ATLAS Collaboration}",
    title          = "{Jet energy scale measurements and their systematic uncertainties in proton--proton collisions at \(\sqrt{s} = 13\,\text{TeV}\) with the ATLAS detector}",
    journal        = "Phys. Rev. D",
    volume         = "96",
    year           = "2017",
    pages          = "072002",
    doi            = "10.1103/PhysRevD.96.072002",
    reportNumber   = "CERN-EP-2017-038",
    eprint         = "1703.09665",
    archivePrefix  = "arXiv",
    primaryClass   = "hep-ex",
}

@Article{PERF-2016-06,
    author         = "{ATLAS Collaboration}",
    title          = "{Identification and rejection of pile-up jets at high pseudorapidity with the ATLAS detector}",
    journal        = "Eur. Phys. J. C",
    volume         = "77",
    year           = "2017",
    pages          = "580",
    doi            = "10.1140/epjc/s10052-017-5081-5",
    reportNumber   = "CERN-EP-2017-055",
    eprint         = "1705.02211",
    archivePrefix  = "arXiv",
    primaryClass   = "hep-ex",
    related        = "PERF-2016-06-err",
    relatedstring  = "Erratum:",
}

@Article{TRIG-2016-01,
    author         = "{ATLAS Collaboration}",
    title          = "{Performance of the ATLAS trigger system in 2015}",
    journal        = "Eur. Phys. J. C",
    volume         = "77",
    year           = "2017",
    pages          = "317",
    doi            = "10.1140/epjc/s10052-017-4852-3",
    reportNumber   = "CERN-EP-2016-241",
    eprint         = "1611.09661",
    archivePrefix  = "arXiv",
    primaryClass   = "hep-ex",
}

@Article{HIGG-2017-02,
    author         = "{ATLAS Collaboration}",
    title          = "{Evidence for the associated production of the Higgs boson and a top quark pair with the ATLAS detector}",
    journal        = "Phys. Rev. D",
    volume         = "97",
    year           = "2018",
    pages          = "072003",
    doi            = "10.1103/PhysRevD.97.072003",
    reportNumber   = "CERN-EP-2017-281",
    eprint         = "1712.08891",
    archivePrefix  = "arXiv",
    primaryClass   = "hep-ex",
}

@Article{DAPR-2018-01,
    author         = "{ATLAS Collaboration}",
    title          = "{ATLAS data quality operations and performance for 2015--2018 data-taking}",
    journal        = "JINST",
    volume         = "15",
    year           = "2020",
    pages          = "P04003",
    doi            = "10.1088/1748-0221/15/04/P04003",
    reportNumber   = "CERN-EP-2019-207",
    eprint         = "1911.04632",
    archivePrefix  = "arXiv",
    primaryClass   = "physics.ins-det",
}

@Article{EGAM-2018-01,
    author         = "{ATLAS Collaboration}",
    title          = "{Electron and photon performance measurements with the ATLAS detector using the 2015--2017 LHC proton--proton collision data}",
    journal        = "JINST",
    volume         = "14",
    year           = "2019",
    pages          = "P12006",
    doi            = "10.1088/1748-0221/14/12/P12006",
    reportNumber   = "CERN-EP-2019-145",
    eprint         = "1908.00005",
    archivePrefix  = "arXiv",
    primaryClass   = "hep-ex",
}

@Article{FTAG-2018-01,
    author         = "{ATLAS Collaboration}",
    title          = "{ATLAS \(b\)-jet identification performance and efficiency measurement with \(t\bar{t}\) events in \(pp\) collisions at \(\sqrt{s} = 13\,\text{TeV}\)}",
    journal        = "Eur. Phys. J. C",
    volume         = "79",
    year           = "2019",
    pages          = "970",
    doi            = "10.1140/epjc/s10052-019-7450-8",
    reportNumber   = "CERN-EP-2019-132",
    eprint         = "1907.05120",
    archivePrefix  = "arXiv",
    primaryClass   = "hep-ex",
}

@Article{HIGG-2018-32,
    author         = "{ATLAS Collaboration}",
    title          = "{Evidence of off-shell Higgs boson production from \(ZZ\) leptonic decay channels and constraints on its total width with the ATLAS detector}",
    journal        = "Phys. Lett. B",
    volume         = "846",
    year           = "2023",
    pages          = "138223",
    doi            = "10.1016/j.physletb.2023.138223",
    reportNumber   = "CERN-EP-2023-023",
    eprint         = "2304.01532",
    archivePrefix  = "arXiv",
    primaryClass   = "hep-ex",
    related        = "HIGG-2018-32-err",
    relatedstring  = "Erratum:",
}

@Article{MUON-2018-03,
    author         = "{ATLAS Collaboration}",
    title          = "{Muon reconstruction and identification efficiency in ATLAS using the full Run~2 \(pp\) collision data set at \(\sqrt{s} = 13\,\text{TeV}\)}",
    journal        = "Eur. Phys. J. C",
    volume         = "81",
    year           = "2021",
    pages          = "578",
    doi            = "10.1140/epjc/s10052-021-09233-2",
    reportNumber   = "CERN-EP-2020-199",
    eprint         = "2012.00578",
    archivePrefix  = "arXiv",
    primaryClass   = "hep-ex",
}

@Article{TRIG-2018-01,
    author         = "{ATLAS Collaboration}",
    title          = "{Performance of the ATLAS muon triggers in Run~2}",
    journal        = "JINST",
    volume         = "15",
    year           = "2020",
    pages          = "P09015",
    doi            = "10.1088/1748-0221/15/09/p09015",
    reportNumber   = "CERN-EP-2020-031",
    eprint         = "2004.13447",
    archivePrefix  = "arXiv",
    primaryClass   = "physics.ins-det",
}

@Article{TRIG-2018-05,
    author         = "{ATLAS Collaboration}",
    title          = "{Performance of electron and photon triggers in ATLAS during LHC Run~2}",
    journal        = "Eur. Phys. J. C",
    volume         = "80",
    year           = "2020",
    pages          = "47",
    doi            = "10.1140/epjc/s10052-019-7500-2",
    reportNumber   = "CERN-EP-2019-169",
    eprint         = "1909.00761",
    archivePrefix  = "arXiv",
    primaryClass   = "hep-ex",
}

@Article{TRIG-2019-03,
    author         = "{ATLAS Collaboration}",
    title          = "{The ATLAS inner detector trigger performance in \(pp\) collisions at \(13\,\text{TeV}\) during LHC Run~2}",
    journal        = "Eur. Phys. J. C",
    volume         = "82",
    year           = "2022",
    pages          = "206",
    doi            = "10.1140/epjc/s10052-021-09920-0",
    reportNumber   = "CERN-EP-2021-076",
    eprint         = "2107.02485",
    archivePrefix  = "arXiv",
    primaryClass   = "hep-ex",
}

@Article{HIGG-2020-07,
    author         = "{ATLAS Collaboration}",
    title          = "{Measurement of the Higgs boson mass in the \(H \to ZZ^* \to 4\ell\) decay channel using \(139\,\text{fb}^{-1}\) of \(\sqrt{s} = 13\,\text{TeV}\) \(pp\) collisions recorded by the ATLAS detector at the LHC}",
    journal        = "Phys. Lett. B",
    volume         = "843",
    year           = "2023",
    pages          = "137880",
    doi            = "10.1016/j.physletb.2023.137880",
    reportNumber   = "CERN-EP-2022-135",
    eprint         = "2207.00320",
    archivePrefix  = "arXiv",
    primaryClass   = "hep-ex",
}

@Article{HIGG-2020-20,
    author         = "{ATLAS Collaboration}",
    title          = "{Measurements of \(WH\) and \(ZH\) production with Higgs boson decays into bottom quarks and direct constraints on the charm Yukawa coupling in \(13\,\text{TeV}\) \(pp\) collisions with the ATLAS detector}",
    year           = "2025",
    reportNumber   = "CERN-EP-2024-237",
    eprint         = "2410.19611",
    archivePrefix  = "arXiv",
    primaryClass   = "hep-ex",
    doi = "10.1007/JHEP04(2025)075",
    journal = "JHEP",
    volume = "04",
    pages = "075",
}

@Article{JETM-2020-03,
    author         = "{ATLAS Collaboration}",
    title          = "{The performance of missing transverse momentum reconstruction and its significance with the ATLAS detector using \(140\,\text{fb}^{-1}\) of \(\sqrt{s} = 13\,\text{TeV}\) \(pp\) collisions}",
    journal        = "Eur. Phys. J. C",
    volume         = "85",
    year           = "2025",
    pages          = "606",
    doi            = "10.1140/epjc/s10052-025-14062-8",
    reportNumber   = "CERN-EP-2024-023",
    eprint         = "2402.05858",
    archivePrefix  = "arXiv",
    primaryClass   = "hep-ex",
}

@Article{DAPR-2021-01,
    author         = "{ATLAS Collaboration}",
    title          = "{Luminosity determination in \(pp\) collisions at \(\sqrt{s} = 13\,\text{TeV}\) using the ATLAS detector at the LHC}",
    journal        = "Eur. Phys. J. C",
    volume         = "83",
    year           = "2023",
    pages          = "982",
    doi            = "10.1140/epjc/s10052-023-11747-w",
    reportNumber   = "CERN-EP-2022-281",
    eprint         = "2212.09379",
    archivePrefix  = "arXiv",
    primaryClass   = "hep-ex",
}

@Article{EGAM-2021-01,
    author         = "{ATLAS Collaboration}",
    title          = "{Electron and photon efficiencies in LHC Run~2 with the ATLAS experiment}",
    journal        = "JHEP",
    volume         = "05",
    year           = "2024",
    pages          = "162",
    doi            = "10.1007/JHEP05(2024)162",
    reportNumber   = "CERN-EP-2023-182",
    eprint         = "2308.13362",
    archivePrefix  = "arXiv",
    primaryClass   = "hep-ex",
}

@Article{SOFT-2022-02,
    author         = "{ATLAS Collaboration}",
    title          = "{Software and computing for Run~3 of the ATLAS experiment at the LHC}",
    journal        = "Eur. Phys. J. C",
    volume         = "85",
    year           = "2025",
    pages          = "234",
    doi            = "10.1140/epjc/s10052-024-13701-w",
    reportNumber   = "CERN-EP-2024-100",
    eprint         = "2404.06335",
    archivePrefix  = "arXiv",
    primaryClass   = "hep-ex",
}

@Article{TOPQ-2023-22,
    author         = "{ATLAS Collaboration}",
    title          = "{Constraint on the total width of the Higgs boson from Higgs boson and four-top-quark measurements in \(pp\) collisions at \(\sqrt{s} = 13\,\text{TeV}\) with the ATLAS detector}",
    journal        = "Phys. Lett. B",
    volume         = "861",
    year           = "2025",
    pages          = "139277",
    doi            = "10.1016/j.physletb.2025.139277",
    reportNumber   = "CERN-EP-2024-190",
    eprint         = "2407.10631",
    archivePrefix  = "arXiv",
    primaryClass   = "hep-ex",
}

@Article{HIGP-2024-07,
    author         = "{ATLAS Collaboration}",
    title          = "{Measurements of Higgs boson production via gluon--gluon fusion and vector-boson fusion using \(H \rightarrow WW^{*} \rightarrow \ell\nu\ell\nu\) decays in \(pp\) collisions with the ATLAS detector and their effective field theory interpretations}",
    year           = "2025",
    reportNumber   = "CERN-EP-2025-054",
    eprint         = "2504.07686",
    archivePrefix  = "arXiv",
    primaryClass   = "hep-ex",
}

@Article{HIGP-2024-14,
    author         = "{ATLAS Collaboration}",
    title          = "{Measurement of off-shell Higgs boson production in the \(H^* \rightarrow ZZ \rightarrow 4\ell\) decay channel using a neural simulation-based inference technique in \(13\,\text{TeV}\) \(pp\) collisions with the ATLAS detector}",
    journal        = "Rept. Prog. Phys.",
    volume         = "88",
    year           = "2025",
    pages          = "057803",
    doi            = "10.1088/1361-6633/adcd9a",
    reportNumber   = "CERN-EP-2024-298",
    eprint         = "2412.01548",
    archivePrefix  = "arXiv",
    primaryClass   = "hep-ex",
}

@Booklet{ATL-SOFT-PUB-2025-001,
    author         = "{ATLAS Collaboration}",
    title          = "{ATLAS Computing Acknowledgements}",
    howpublished   = "{ATL-SOFT-PUB-2025-001}",
    url            = "https://cds.cern.ch/record/2922210",
    year           = "2025",
}

@Article{LUCID2,
    author         = "Avoni, G. and others",
    title          = "{The new LUCID-2 detector for luminosity measurement and monitoring in ATLAS}",
    journal        = "JINST",
    volume         = "13",
    year           = "2018",
    pages          = "P07017",
    doi            = "10.1088/1748-0221/13/07/P07017",
}

@Article{CMS-HIG-12-028,
    author         = "{CMS Collaboration}",
    title          = "{Observation of a new boson at a mass of 125 GeV with the CMS experiment at the LHC}",
    journal        = "Phys. Lett. B",
    volume         = "716",
    year           = "2012",
    pages          = "30",
    doi            = "10.1016/j.physletb.2012.08.021",
    reportNumber   = "CERN-PH-EP-2012-220",
    eprint         = "1207.7235",
    archivePrefix  = "arXiv",
    primaryClass   = "hep-ex",
}

@Article{CMS-HIG-14-032,
    author         = "{CMS Collaboration}",
    title          = "{Search for Higgs boson off-shell production in proton--proton collisions at 7 and \(8\,\text{TeV}\) and derivation of constraints on its total decay width}",
    journal        = "JHEP",
    volume         = "09",
    year           = "2016",
    pages          = "051",
    doi            = "10.1007/JHEP09(2016)051",
    reportNumber   = "CERN-EP-2016-054",
    eprint         = "1605.02329",
    archivePrefix  = "arXiv",
    primaryClass   = "hep-ex",
}

@Article{CMS-HIG-21-013,
    author         = "{CMS Collaboration}",
    title          = "{Measurement of the Higgs boson width and evidence of its off-shell contributions to \(ZZ\) production}",
    journal        = "Nature Phys.",
    volume         = "18",
    year           = "2022",
    pages          = "1329--1334",
    doi            = "10.1038/s41567-022-01682-0",
    reportNumber   = "CERN-EP-2021-272",
    eprint         = "2202.06923",
    archivePrefix  = "arXiv",
    primaryClass   = "hep-ex",
}

@Article{CMS-HIG-21-019,
    author         = "{CMS Collaboration}",
    title          = "{Measurement of the Higgs boson mass and width using the four-lepton final state in proton--proton collisions at \(\sqrt{s} = 13\,\text{TeV}\)}",
    journal        = "Phys. Rev. D",
    volume         = "111",
    year           = "2024",
    pages          = "092014",
    doi            = "10.1103/PhysRevD.111.092014",
    reportNumber   = "CERN-EP-2024-210",
    eprint         = "2409.13663",
    archivePrefix  = "arXiv",
    primaryClass   = "hep-ex",
}

@Booklet{ATLAS-CONF-2015-029,
    author         = "{ATLAS Collaboration}",
    title          = "{Selection of jets produced in \(13~\text{TeV}\) proton--proton collisions with the ATLAS detector}",
    howpublished   = "{ATLAS-CONF-2015-029}",
    url            = "https://cds.cern.ch/record/2037702",
    year           = "2015",
}

@article{Alioli:2021,
  title={Four-lepton production in gluon fusion at NLO matched to parton showers},
  author={Alioli, Simone and Ravasio, Silvia Ferrario and Lindert, Jonas M and R{\"o}ntsch, Raoul},
  journal = "Eur. Phys. J. C",
  volume={81},
  number={8},
  doi={10.1140/epjc/s10052-021-09470-5},
  pages={687},
  year={2021},
  publisher={Springer},
  eprint         = "2102.07783",
  archivePrefix  = "arXiv",
  primaryClass   = "hep-ph",
}

@article{Plehn:2000,
  title={A Method for identifying $H\to\tau\tau \to e^{\pm} \mu^{\mp} /\!\!\!{p_T}$ at the CERN LHC},
  author={Plehn, T and Rainwater, D and Zeppenfield, D},
  journal="Phys. Rev. D",
  volume={61},
  number={9},
  pages={093005},
  year={2000},
  publisher={APS},
  doi={10.1103/PhysRevD.61.093005},
  eprint         = "hep-ph/9911385",
  archivePrefix  = "arXiv",
}

@article{Campbell:2014,
  title = {Bounding the Higgs width at the LHC: Complementary results from $H\ensuremath{\rightarrow}WW$},
  author = {Campbell, John M. and Ellis, R. Keith and Williams, Ciaran},
  journal = {Phys. Rev. D},
  volume = {89},
  pages = {053011},
  numpages = {7},
  year = {2014},
  month = {3},
  publisher = {American Physical Society},
  doi = {10.1103/PhysRevD.89.053011},
  eprint         = "1312.1628",
  archivePrefix  = "arXiv",
  primaryClass   = "hep-ph",
}

@article{Caola:2013,
  title = {Constraining the Higgs boson width with $ZZ$ production at the LHC},
  author = {Caola, Fabrizio and Melnikov, Kirill},
  journal = {Phys. Rev. D},
  volume = {88},
  pages = {054024},
  year = {2013},
  month = {9},
  publisher = {American Physical Society},
  doi = {10.1103/PhysRevD.88.054024},
  eprint         = "1307.4935",
  archivePrefix  = "arXiv",
  primaryClass   = "hep-ph",
}

@article{Cao:2023,
doi = {10.1088/1674-1137/aca8f6},
year = {2023},
month = {3},
publisher = {Chinese Physical Society and the Institute of High Energy Physics of the Chinese Academy of Sciences and the Institute of Modern Physics of the Chinese Academy of Sciences and IOP Publishing Ltd},
eprint         = "2107.08343",
primaryClass   = "hep-ph",
archivePrefix  = "arXiv",
volume = {47},
number = {3},
pages = {033101},
author = {Qing-Hong Cao and Hao-Lin Li and Ling-Xiao Xu and Jiang-Hao Yu},
title = {What can we learn from the total width of the Higgs boson?},
journal = {Chin. Phys. C}
}

@article{wilks1938large,
  title={The large-sample distribution of the likelihood ratio for testing composite hypotheses},
  author={Wilks, Samuel S},
  journal={Ann. Math. Statist.},
  volume={9},
  number={1},
  pages={60--62},
  year={1938},
  doi = "10.1214/aoms/1177732360",
  %URL = {http://www.jstor.org/stable/2957648},
  publisher={JSTOR}
}

@article{neyman1937outline,
  title={Outline of a Theory of Statistical Estimation Based on the Classical Theory of Probability},
  author={Neyman, Jerzy},
  journal={Phil. Trans. Roy. Soc. Lond. A},
  volume={236},
  number={767},
  year={1937},
  pages={767},
  doi = {10.1098/rsta.1937.0005},
  %url = {https://api.semanticscholar.org/CorpusID:19584450},
  publisher={The Royal Society London}
}

@article{agashe2005minimal,
  title={The minimal composite Higgs model},
  author={Agashe, Kaustubh and Contino, Roberto and Pomarol, Alex},
  journal={Nucl. Phys. B},
  volume={719},
  number={1-2},
  pages={165--187},
  year={2005},
  %url={http://dx.doi.org/10.1016/j.nuclphysb.2005.04.035},
  DOI={10.1016/j.nuclphysb.2005.04.035},
  eprint         = "0412089",
  archivePrefix  = "arXiv",
  primaryClass   = "hep-ph",
  publisher={Elsevier}
}

@article{branco2012theory,
  title={Theory and phenomenology of two-Higgs-doublet models},
  author={Branco, Gustavo Castelo and Ferreira, PM and Lavoura, L and Rebelo, MN and Sher, Marc and Silva, Joao P},
  journal={Phys. Rept.},
  volume={516},
  number={1-2},
  pages={1--102},
  year={2012},
  %url={http://dx.doi.org/10.1016/j.physrep.2012.02.002},
  DOI={10.1016/j.physrep.2012.02.002},
  eprint         = "1106.0034",
  archivePrefix  = "arXiv",
  primaryClass   = "hep-ph",
  publisher={Elsevier}
}

@article{alonso2016sigma,
  title={Sigma models with negative curvature},
  author={Alonso, Rodrigo and Jenkins, Elizabeth E and Manohar, Aneesh V},
  journal={Phys. Lett. B},
  volume={756},
  pages={358--364},
  year={2016},
  %url={http://dx.doi.org/10.1016/j.physletb.2016.03.032},
  DOI={10.1016/j.physletb.2016.03.032},
  eprint         = "1602.00706",
  archivePrefix  = "arXiv",
  primaryClass   = "hep-ph",
  publisher={Elsevier}
}

@article{georgi1985doubly,
  title={Doubly charged Higgs bosons},
  author={Georgi, Howard and Machacek, Marie},
  journal={Nucl. Phys. B},
  volume={262},
  number={3},
  pages={463--477},
  year={1985},
  doi = {https://doi.org/10.1016/0550-3213(85)90325-6},
  %url = {https://www.sciencedirect.com/science/article/pii/0550321385903256},
  publisher={Elsevier}
}

@Booklet{ATL-PHYS-PUB-2014-021,
    author         = "{ATLAS Collaboration}",
    title          = "{ATLAS Pythia~8 tunes to \(7~\text{TeV}\) data}",
    howpublished   = "{ATL-PHYS-PUB-2014-021}",
    url            = "https://cds.cern.ch/record/1966419",
    year           = "2014",
}

@Booklet{ATL-PHYS-PUB-2016-017,
    author         = "{ATLAS Collaboration}",
    title          = "{The Pythia~8 A3 tune description of ATLAS minimum bias and inelastic measurements incorporating the Donnachie--Landshoff diffractive model}",
    howpublished   = "{ATL-PHYS-PUB-2016-017}",
    url            = "https://cds.cern.ch/record/2206965",
    year           = "2016",
}

@Booklet{ATL-PHYS-PUB-2016-020,
    author         = "{ATLAS Collaboration}",
    title          = "{Studies on top-quark Monte Carlo modelling for Top2016}",
    howpublished   = "{ATL-PHYS-PUB-2016-020}",
    url            = "https://cds.cern.ch/record/2216168",
    year           = "2016",
}

@Booklet{ATL-PHYS-PUB-2019-026,
    author         = "{ATLAS Collaboration}",
    title          = "{Forward jet vertex tagging using the particle flow algorithm}",
    howpublished   = "{ATL-PHYS-PUB-2019-026}",
    url            = "https://cds.cern.ch/record/2683100",
    year           = "2019",
}

@Booklet{ATL-PHYS-PUB-2021-042,
    author         = "{ATLAS Collaboration}",
    title          = "{Studies of \(t\bar{t} / tW\) interference effects in \(b\bar{b}\ell^{+}\ell^{-'}\nu\bar{\nu}'\) final states with \textsc{Powheg} and \textsc{MadGraph5\_aMC@NLO} setups}",
    howpublished   = "{ATL-PHYS-PUB-2021-042}",
    url            = "https://cds.cern.ch/record/2792254",
    year           = "2021",
}
\clearpage
% ATLAS Collaboration author list
% Reference date of HIGP-2024-05 is 2025-02-14
% Author list last updated on date 24-JUL-25
% Data extracted on 24-Jul-2025 for paper reference HIGP-2024-05
% at 4:24pm
 
\begin{flushleft}
\hypersetup{urlcolor=black}
{\Large The ATLAS Collaboration}

\bigskip

\AtlasOrcid[0000-0002-6665-4934]{G.~Aad}$^\textrm{\scriptsize 104}$,
\AtlasOrcid[0000-0001-7616-1554]{E.~Aakvaag}$^\textrm{\scriptsize 17}$,
\AtlasOrcid[0000-0002-5888-2734]{B.~Abbott}$^\textrm{\scriptsize 123}$,
\AtlasOrcid[0000-0002-0287-5869]{S.~Abdelhameed}$^\textrm{\scriptsize 119a}$,
\AtlasOrcid[0000-0002-1002-1652]{K.~Abeling}$^\textrm{\scriptsize 55}$,
\AtlasOrcid[0000-0001-5763-2760]{N.J.~Abicht}$^\textrm{\scriptsize 49}$,
\AtlasOrcid[0000-0002-8496-9294]{S.H.~Abidi}$^\textrm{\scriptsize 30}$,
\AtlasOrcid[0009-0003-6578-220X]{M.~Aboelela}$^\textrm{\scriptsize 45}$,
\AtlasOrcid[0000-0002-9987-2292]{A.~Aboulhorma}$^\textrm{\scriptsize 36e}$,
\AtlasOrcid[0000-0001-5329-6640]{H.~Abramowicz}$^\textrm{\scriptsize 157}$
% ...
\bigskip
\\

% ...
$^{17}$Department for Physics and Technology, University of Bergen, Bergen; Norway.\\
% ...
$^{30}$Physics Department, Brookhaven National Laboratory, Upton NY; United States of America.\\
% ...
$^{36}$$^{(a)}$Facult\'e des Sciences Ain Chock, Universit\'e Hassan II de Casablanca;$^{(b)}$Facult\'{e} des Sciences, Universit\'{e} Ibn-Tofail, K\'{e}nitra;$^{(c)}$Facult\'e des Sciences Semlalia, Universit\'e Cadi Ayyad, LPHEA-Marrakech;$^{(d)}$LPMR, Facult\'e des Sciences, Universit\'e Mohamed Premier, Oujda;$^{(e)}$Facult\'e des sciences, Universit\'e Mohammed V, Rabat;$^{(f)}$Institute of Applied Physics, Mohammed VI Polytechnic University, Ben Guerir; Morocco.\\
% ...
$^{45}$Physics Department, Southern Methodist University, Dallas TX; United States of America.\\
% ...
$^{49}$Fakult\"{a}t Physik , Technische Universit{\"a}t Dortmund, Dortmund; Germany.\\
% ...
$^{55}$II. Physikalisches Institut, Georg-August-Universit\"{a}t G\"ottingen, G\"ottingen; Germany.\\
% ...
$^{104}$CPPM, Aix-Marseille Universit\'e, CNRS/IN2P3, Marseille; France.\\
% ...
$^{119}$$^{(a)}$New York University Abu Dhabi, Abu Dhabi;$^{(b)}$United Arab Emirates University, Al Ain; United Arab Emirates.\\
% ...
$^{123}$Homer L. Dodge Department of Physics and Astronomy, University of Oklahoma, Norman OK; United States of America.\\
% ...
$^{157}$Raymond and Beverly Sackler School of Physics and Astronomy, Tel Aviv University, Tel Aviv; Israel.\\
% ...
$^{a}$ Also at Affiliated with an institute formerly covered by a cooperation agreement with CERN.\\
$^{b}$ Also at An-Najah National University, Nablus; Palestine.\\
$^{c}$ Also at Borough of Manhattan Community College, City University of New York, New York NY; United States of America.\\
$^{d}$ Also at Center for Interdisciplinary Research and Innovation (CIRI-AUTH), Thessaloniki; Greece.\\
$^{e}$ Also at Centre of Physics of the Universities of Minho and Porto (CF-UM-UP); Portugal.\\
$^{f}$ Also at CERN, Geneva; Switzerland.\\
% ...

\end{flushleft}

% Created with Glance <Atlas.Glance@cern.ch>

%

%
%

%

%
%
%
%

\end{document}